\documentclass[letter,11pt]{article}
\usepackage{float}
\usepackage{hyperref}
\usepackage{graphicx, amssymb, wrapfig,setspace,multicol,natbib,hyperref,setspace}
\def\ltsima{$\; \buildrel < \over \sim \;$}
\def\simlt{\lower.5ex\hbox{\ltsima}}
\def\gtsima{$\; \buildrel > \over \sim \;$}
\def\simgt{\lower.5ex\hbox{\gtsima}}
\newcommand{\uJy}{$\mu$Jy}
\newcommand{\araa}{ARA\&A}
\newcommand{\apj}{ApJ}
\newcommand{\physrep}{Physics Reports}
\newcommand{\apjl}{ApJL}
\newcommand{\jqsrt}{JQSRT}
\newcommand{\mnras}{MNRAS}
\newcommand{\aap}{A\&A}

\newcommand{\apjs}{ApJS}

\newcommand{\nat}{{\it Nature}}
\newcommand{\etal}{et~al.}
\newcommand{\lsun}{L$_{\odot}$}
\newcommand{\msun}{M$_{\odot}$}

\begin{document}

\begin{center}
\includegraphics[width=\textwidth]{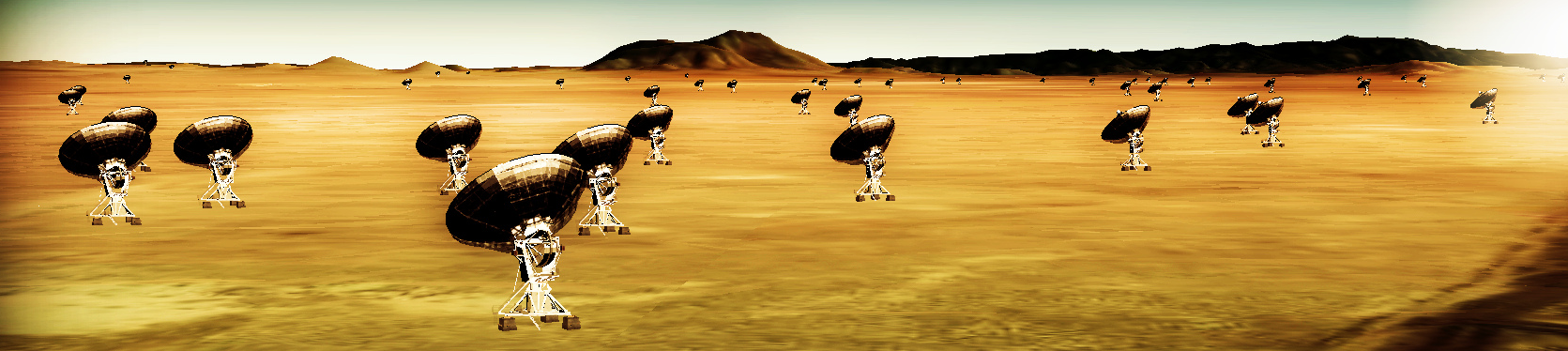}
\end{center}

\begin{center}

{\bf \Large Next Generation Very Large Array Memo No. 8}

\vspace{0.1in}

{\bf \Large Science Working Group 3}

\vspace{0.1in}

{\bf \Large Galaxy Assembly through Cosmic Time}

\end{center}

\hrule

\vspace{0.7cm}

\noindent Caitlin M. Casey$^{1}$,
Jacqueline A. Hodge$^{2,3}$,
Mark Lacy$^{2}$,
Christopher A. Hales$^{4}$,
Amy Barger$^{5}$,
Desika Narayanan$^{6}$,
Chris Carilli$^{4,7}$,
Katherine Alatalo$^{8}$,
Elisabete da Cunha$^{9}$,
Bjorn Emonts$^{10}$,
Rob Ivison$^{11,12}$,
Amy Kimball$^{13}$,
Kotaro Kohno$^{14}$,
Eric Murphy$^{15}$,
Dominik Riechers$^{16}$,
Mark Sargent$^{17}$,
Fabian Walter$^{18}$\\

\vspace{0.7cm}

\begin{center}
{\bf \large Abstract}\\
\end{center}

The Next-Generation Very Large Array (ngVLA) will be critical for
understanding how galaxies are built and evolve at the earliest
epochs.  The sensitivity and frequency coverage will allow for the
detection of cold gas and dust in `normal' distant galaxies, including
the low-J transitions of molecular gas tracers such as CO, HNC, HCN,
and HCO+; synchrotron and free-free continuum emission; and even the
exciting possibility of thermal dust emission at the highest
($z\sim7$) redshifts.  In particular, by enabling the total molecular
gas reservoirs to be traced to unprecedented sensitivities across a
huge range of epochs simultaneously -- something no other radio or
submillimeter facility will be capable of -- the detection of the
crucial low-J transitions of CO in a diverse body of galaxies will be
the cornerstone of ngVLA's contribution to high-redshift galaxy
evolution science.  The ultra-wide bandwidths will allow a complete
sampling of radio SEDs, as well as the detection of emission lines
necessary for spectroscopic confirmation of elusive dusty starbursts.
The ngVLA will also deliver unique contributions to our understanding
of cosmic magnetism and to science accessible through microwave
polarimetry.  Finally, the superb angular resolution will move the
field beyond detection experiments and allow detailed studies of the
morphology and dynamics of these systems, including dynamical modeling
of disks/mergers, determining the properties of outflows, measuring
black hole masses from gas disks, and resolving multiple AGN nuclei.
We explore the contribution of a ngVLA to these areas and more, as
well as synergies with current and upcoming facilities including ALMA,
SKA, large single-dish submillimeter observatories, GMT/TMT, and JWST.

\vspace{2mm}

\begin{spacing}{1}
{\footnotesize
\noindent $^{1}$Department of Astronomy, The University of Texas at Austin, 2515 Speedway Blvd, Austin, TX 78712, USA\\
\noindent $^{2}$National Radio Astronomy Observatory, 520 Edgemont Road, Charlottesville, VA 22903, USA\\
\noindent $^{3}$Leiden University, J.H. Oort Building, Niels Bohrweg 2 NL-2333 CA Leiden, The Netherlands\\
\noindent $^{4}$National Radio Astronomy Observatory, 1003 Lopezville Rd, Socorro, NM 87801, USA\\
\noindent $^{5}$Department of Astronomy, University of Wisconsin--Madison, 475 North Charter Street, Madison, WI 53706, USA\\
\noindent $^{6}$Department of Physics \&\ Astronomy, Haverford College, 370 Lancaster Ave, Haverford, PA 19041, USA\\
\noindent $^{7}$Cavendish Astrophysics Group, JJ Thomson Avenue, Cambridge, CB3 0HE, UK\\
\noindent $^{8}$Carnegie Observatories, 813 Santa Barbara St, Pasadena, CA 91101, USA\\
\noindent $^{9}$Center for Astrophysics and Supercomputing, Swinburne University of Technology, Hawthorn VIC 3122, Australia\\
\noindent $^{10}$Centro de Astrobiologia, Instituto Nacional de T\'{e}cnica Aeroespacial, 28850 Torrej\'{o}n de Ardoz, Madrid, Spain\\
\noindent $^{11}$European Southern Observatory, Karl-Schwarzschild-Strasse 2, D-85748 Garching, Germany\\
\noindent $^{12}$Institute for Astronomy, University of Edinburgh, Royal Observatory,
Blackford Hill, Edinburgh EH9 3HJ, UK\\
\noindent $^{13}$CSIRO Astronomy \& Space Science, Epping, PO Box 76, NSW 1710 Australia\\
\noindent $^{14}$Institute of Astronomy, The University of Tokyo,
2-21-1 Osawa, Mitaka, Tokyo, 181-0015 Japan\\
\noindent $^{15}$IPAC, Caltech, MC 220-6, Pasadena, CA 91125, USA\\
\noindent $^{16}$Astronomy Department, Cornell University,
220 Space Sciences Building, Ithaca, NY 14853, USA\\
\noindent $^{17}$CEA Saclay, DSM/Irfu/Service dÕAstrophysique, Orme des Merisiers, F-91191 Gif-sur-Yvette Cedex, France\\
\noindent $^{18}$Max-Planck-Institut f\"ur Astronomie, K\"onigstuhl 17, D-69117 Heidelberg, Germany\\
}
\end{spacing}

\tableofcontents

\newpage

\section{Introduction}


The Karl G. Jansky Very Large Array (VLA) is a 27-element radio
interferometer located on the Plains of San Agustin, New Mexico, which
has enabled cutting-edge radio astronomy for over three decades. Each
of the 25-m diameter antennae are situated on rail tracks that form
the shape of a `Y', providing baselines up to 36 km and angular
resolutions as high as $\sim$0.04$''$ at the top end of the 74--50,000
MHz frequency range.  Originally known as simply the Very Large Array,
the array was renamed as the Jansky Very Large Array in 2012 to mark
the completion of a major expansion of its capabilities. This upgrade
to several of the major hardware areas allowed for complete frequency
coverage between 1--50 GHz, an order of magnitude improvement in
continuum sensitivity, and increased spectroscopic capability and
correlator flexibility.

In late 2014, NRAO established several working groups in order to
explore the scientific potential of a next-generation large array for
centimeter-wave (0.3--30 cm) astronomy, well beyond the current
capabilities of even the JVLA.  Currently designated the `Next
Generation Very Large Array' (ngVLA), such an array would entail ten
times the effective collecting area of the JVLA and ALMA, operating
from 1\,GHz to 115\,GHz, with ten times longer baselines (300\,km)
providing mas-resolution, plus a dense core on km-scales for high
surface brightness imaging \citep{carilli15a}. The ngVLA is envisioned
to represent a factor of 5--10$\times$ improvement over the
present-day VLA in multiple parameters, including frequency coverage,
instantaneous radio frequency (RF) and processed bandwidth, total
collecting area, maximum baseline length, field of view, and
correlator modes. These working groups have subsequently identified a
number of key science goals that would benefit greatly from a
substantial upgrade to one or more of these parameters, and for which
no other existing or upcoming facility will suffice.

The high-redshift science goals of the ngVLA $-$ which are the key
focus of this report (working group 3) $-$ are intimately intertwined
with some of the most important questions surrounding galaxy evolution
studies today. Specifically, how and when did most galaxies in the
Universe assemble? What physical processes dominate the buildup of
stellar mass in the Universe, and how has that changed and evolved
over the past 12-13 billion years?  What role does environment play in
galaxy growth?  Can our understanding of the first galaxies shed light
on early Universe cosmology, including hierarchical growth and
collapse of large scale structure?
In this white paper, we explore the many important contributions that
the ngVLA will make to studies of galaxy assembly over cosmic time and
out to the earliest epochs of galaxy formation.

\section{Motivation}

\begin{figure}
\vspace*{-1.3cm}
\centering
\includegraphics[natwidth=5in,width=5in]{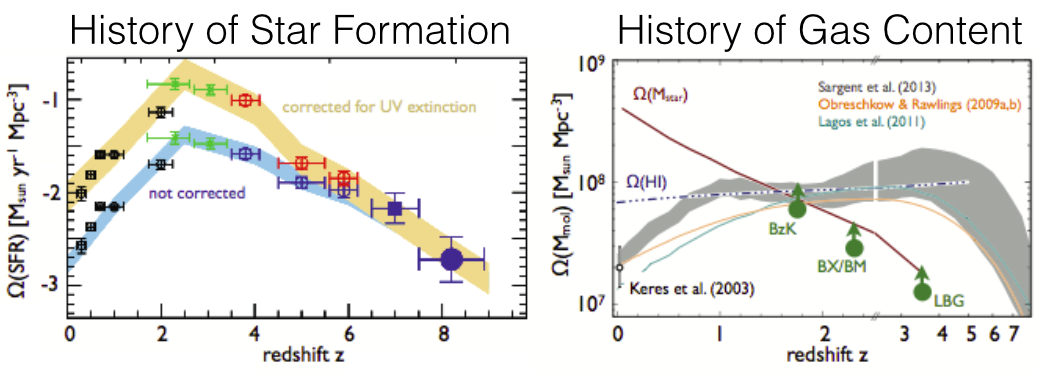}
\begin{spacing}{0.7}
\caption{\footnotesize At left we show the total star formation rate density 
(SFRD) of the universe as a function of redshift as traced by deep
UV/optical surveys out to $z\sim8$ \citep{bouwens09a}. The two curves
indicate the SFRD measured directly from star-light (blue) and after
correction for estimated dust content (tan), where the latter is
extrapolated from the rest-frame ultraviolet slope of galaxies. Note
that direct dust observations of the SFRD currently only extend to
$z\sim2-4$.  While much work has been done on the history of star
formation, there has been relatively little work done on the history
of the Universe's gas content, shown at right.  Observational
constraints on the molecular gas content of galaxies $-$ the fuel for
star formation $-$ are far from complete, and models are not in clear
agreement as to what the redshift evolution of the gas content should
resemble.  Figure adapted from \citet{carilli13a}. }
\end{spacing}
\label{fig:sfrd}
\end{figure}

Deep, legacy optical surveys of the Universe have laid down a
framework for our understanding of the formation of galaxies at early
times. They have indicated that galaxies were forming stars at rates
hundreds of times higher at $z\sim2$ than in the present day.  This is
the epoch where the star formation rate density (SFRD) peaks, and at
earlier times, very deep surveys attempt to learn when galaxies first
became illuminated (see Figure~\ref{fig:sfrd} for a depiction of the
redshift evolution of the SFRD from optical surveys).
However, at redshifts beyond $z\sim1.5$, the measurements which constrain this understanding have
largely been limited to optical and near-infrared pencil-beam surveys
of direct starlight, thus limited in dynamic range, accessible volume,
and to galaxies readily visible via their direct emission from stars.
This, despite the fact that half of all energy emitted by
extragalactic sources is obscured and emitted by dust or gas at much
longer wavelengths.

The submillimeter and radio long-wavelength regime tell a crucial part
of the galaxy evolution story, but they have been largely set aside by
a large fraction of the galaxy evolution community in past decades due
to limitations in long-wavelength instrumentation and sensitivity.
Yet recent years have seen a dramatic increase in our understanding of
galaxy evolution from a bolometric point of view, from both dust and
gas emission, thanks to facilities like the Very Large Array (VLA),
{\sc Scuba} on the JCMT, the {\it Herschel Space Observatory}, IRAM's
Plateau de Bure Interferometer (PdBI), and the Atacama Large
Millimeter Array (ALMA).  Here we present the argument that the next
few decades will see a significant shift in galaxy evolution work away
from the rest-frame optical/UV and towards long-wavelength studies.
The ngVLA will play a critical part in defining what we will be able
to learn about galaxy assembly and large scale structure.

Cold molecular gas is the most fundamental building block for star
formation, yet the Universe's molecular gas content is very poorly
understood relative to its measured stellar content
(Figure~\ref{fig:sfrd}).  Solving the existing mysteries of how the
Universe's stars assembled requires a detailed look at such a
fundamental building block. Cosmological simulations are now making
detailed predictions as to how this gas migrates, heats, cools, and
enters and leaves galaxies via accretion and feedback mechanisms.
While these simulations represent a huge leap forward in theoretical
modeling, these predictions have yet to be tested through direct
observation of the molecular gas at high-redshift.

In addition to the much needed molecular gas census at high-redshift,
the galaxy evolution community also needs a more detailed understanding of
how galaxies in the distant Universe may be fundamentally different
physically than those in the low-redshift Universe.  Dynamics play a
key role in the distinction.  While high-redshift galaxies seem to be
substantially different from nearby galaxies in terms of their star formation rates and stellar mass
characteristics \citep{noeske07a,daddi08a}, it is 
unclear whether or not these differences are driven primarily by
cosmic downsizing associated with the different conditions of the
intergalactic medium \citep{cowie86a}, or if the galaxies grow in
fundamentally different ways at high-redshift.  Probing the internal
kinematics through rest-frame millimeter molecular lines gets to the
heart of this debate, but to-date, has only been done for a handful of
galaxies beyond $z>2$.

\section{Cold Gas Emission}

Cold molecular gas is the fuel which powers star formation, the
production of metals, the formation of stellar byproducts, and
therefore galaxy growth.  The bulk of the Universe's cold gas reservoir
is comprised of hydrogen gas in the form of atomic hydrogen,
H{\sc i}, and molecular hydrogen, H$_{2}$, the latter of which is
responsible for star formation in cold, condensed molecular clouds.
Though H$_{2}$ is not directly observable under normal circumstances,
CO gas is typically used as a proxy given that its particularly bright
transitions are tightly correlated with the ambient density of
molecular hydrogen.  This section discusses the primary goals of ngVLA
science as they relate to cold gas at high-redshift, through
observations of CO and tracers of denser gas phases.

\subsection{CO}\label{sec:co}

\begin{figure}
\centering
\includegraphics[width=0.99\textwidth,natwidth=800]{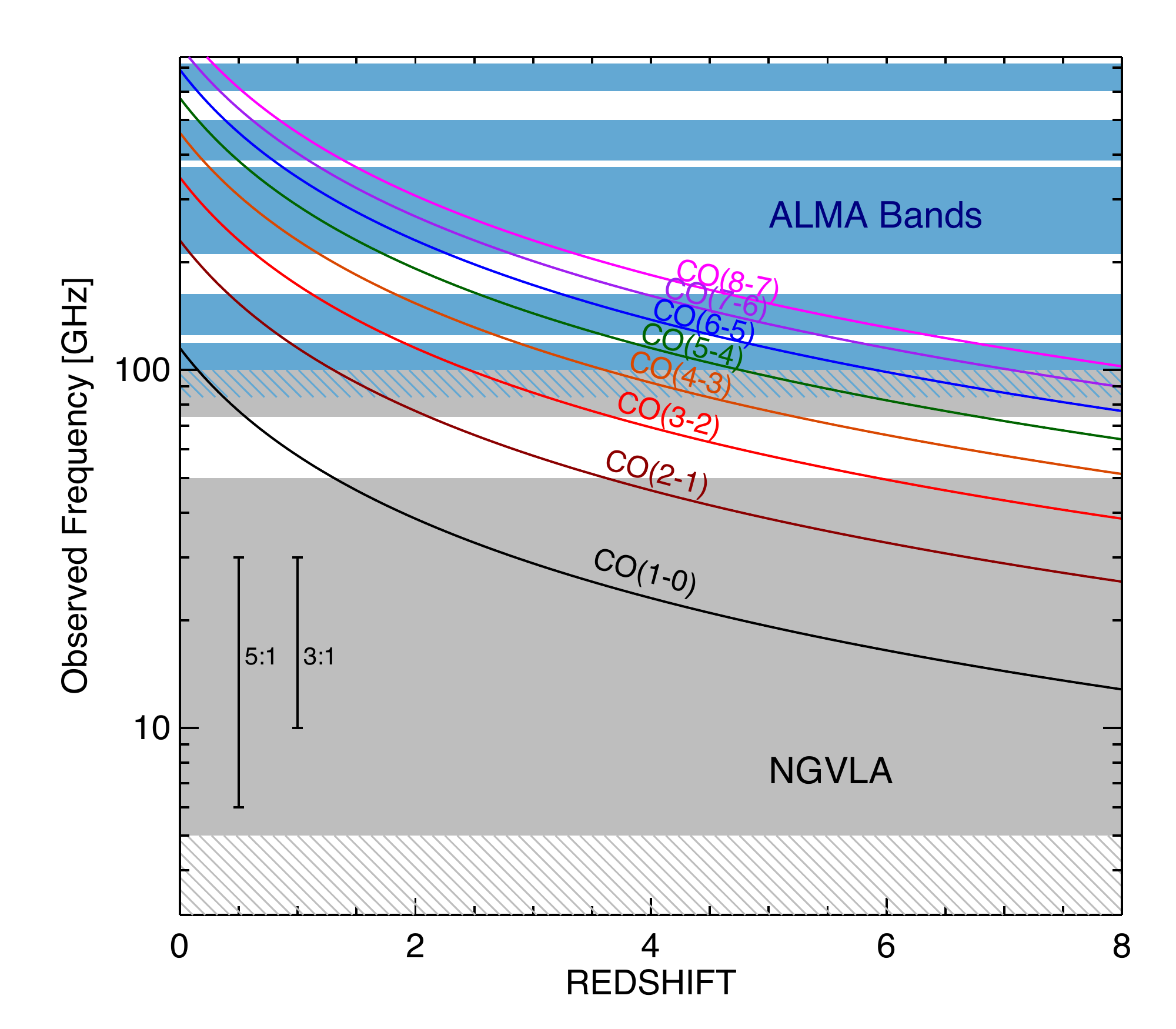}
\begin{spacing}{0.7}
\caption{\footnotesize The observed frequencies of various transitions of CO 
from $z=0$ to $z=8$.  While the ALMA bands are sensitive to many
transitions of CO at low redshift, the low-J CO lines redshift out of
the bands beyond $z\sim0.5-2$.  These low-J CO lines are critical for
characterization of the total molecular gas reservoir in high-$z$
galaxies and are only currently accessible for high-redshift galaxies
using the VLA (lower gray region plus gray hashed). The blue regions
represent ALMA bands, and the blue hashed region represents the
hypothetical ALMA band 0).  The gray region indicates the ngVLA with a
frequency coverage of 5--100\,GHz, with a gap around the 59--75\,GHz
water band.  The gray hashed region represents the reach capabilities
for ngVLA at low frequencies.  With the advent of very wide-bandwidth
setups for the ngVLA, we will be able to obtain multiple CO
transitions for galaxies beyond $z\sim3.5$ in one correlator setup,
which will be revolutionary for studying dust and gas in very
high-redshift galaxies.}
\end{spacing}
\label{fig:coladder}
\end{figure}

Transitions of carbon monoxide are particularly bright due to CO's low
critical density, making it the easiest molecule to detect in
high-redshift galaxies.  Indeed, the vast majority of the few hundred
molecular line detections in the high-redshift Universe have been of
CO \citep[see more in reviews of][]{solomon05a,carilli13a}.  Due to
frequency shifting at sufficiently high-redshift, millimeter-operating
facilities like IRAM's PdBI, CARMA and ALMA can only observe the
higher-J transitions of CO. The lower-J transitions, including the ground state CO(1-0) transition, are only
accessible to longer wavelength radio observatories like the VLA.  See
Figure~\ref{fig:coladder} for an illustration of this problem for
surveying CO (particularly low-J CO) at high-redshift.

While high-J lines are useful for confirming the existence of
galaxies' molecular gas reservoirs, higher-J CO transitions alone
cannot constrain the total molecular gas reservoir, which requires two
significantly uncertain conversions.  The first is the conversion of
high-J CO line strength to the ground state CO(1-0) line
strength. This conversion requires knowledge of internal gas
temperature and excitation, which might vary substantially in
different regions of the galaxy by up to factors of five, and even between
the relatively low-J transitions from CO(3-2) to CO(1-0).  The second
uncertainty is the CO--H$_{2}$ conversion factor, X$_{\rm CO}$ or
$\alpha_{\rm CO}$, which is required to convert from CO(1-0) to the mass of molecular hydrogen. 
This conversion varies by up to a factor of five depending on
the conditions of the gas in the ISM, including metallicity.

\begin{figure}
\centering
\includegraphics[width=0.70\textwidth,natwidth=612,natheight=792]{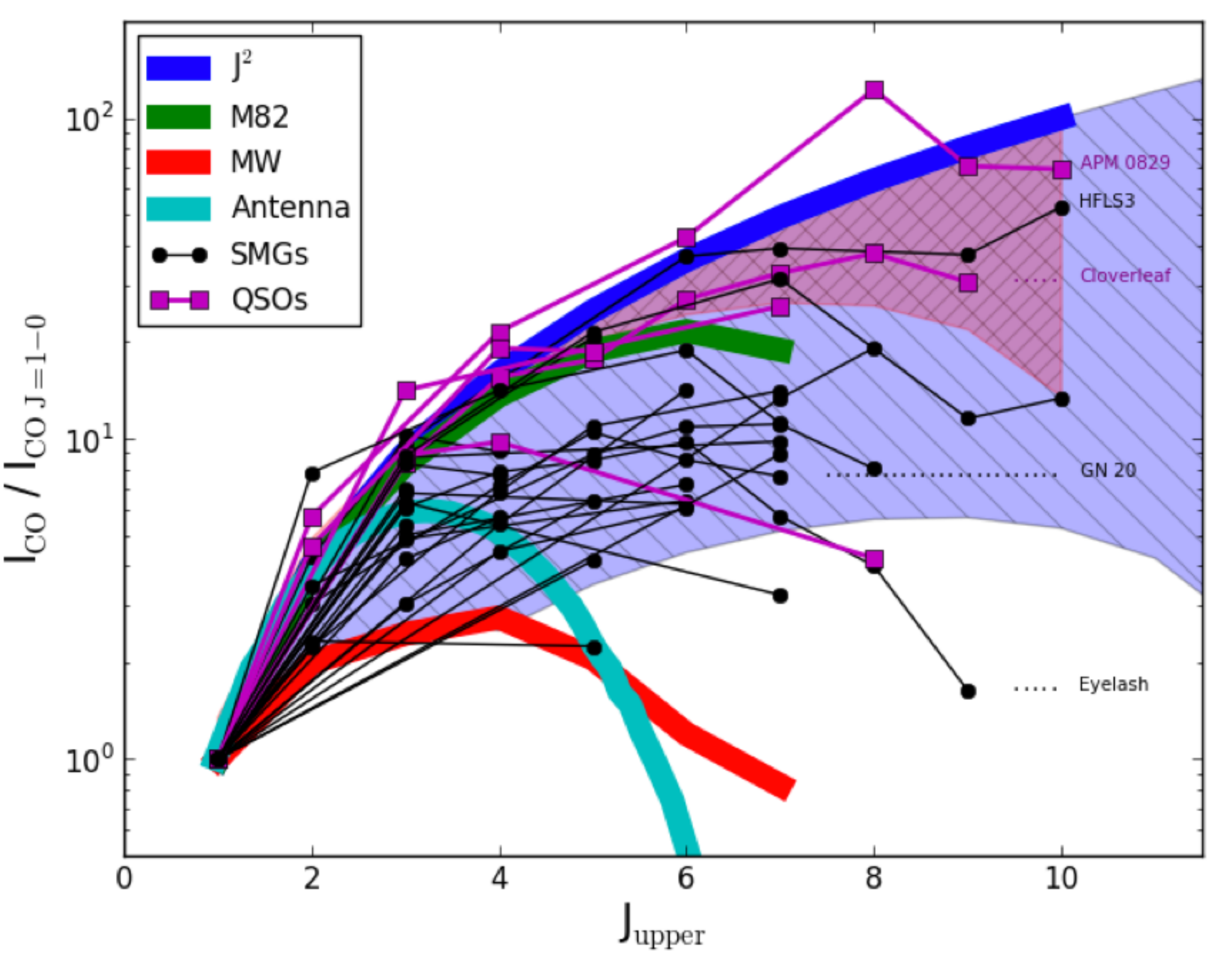}
\hspace{3mm}\includegraphics[width=0.75\textwidth,natwidth=612,natheight=792]{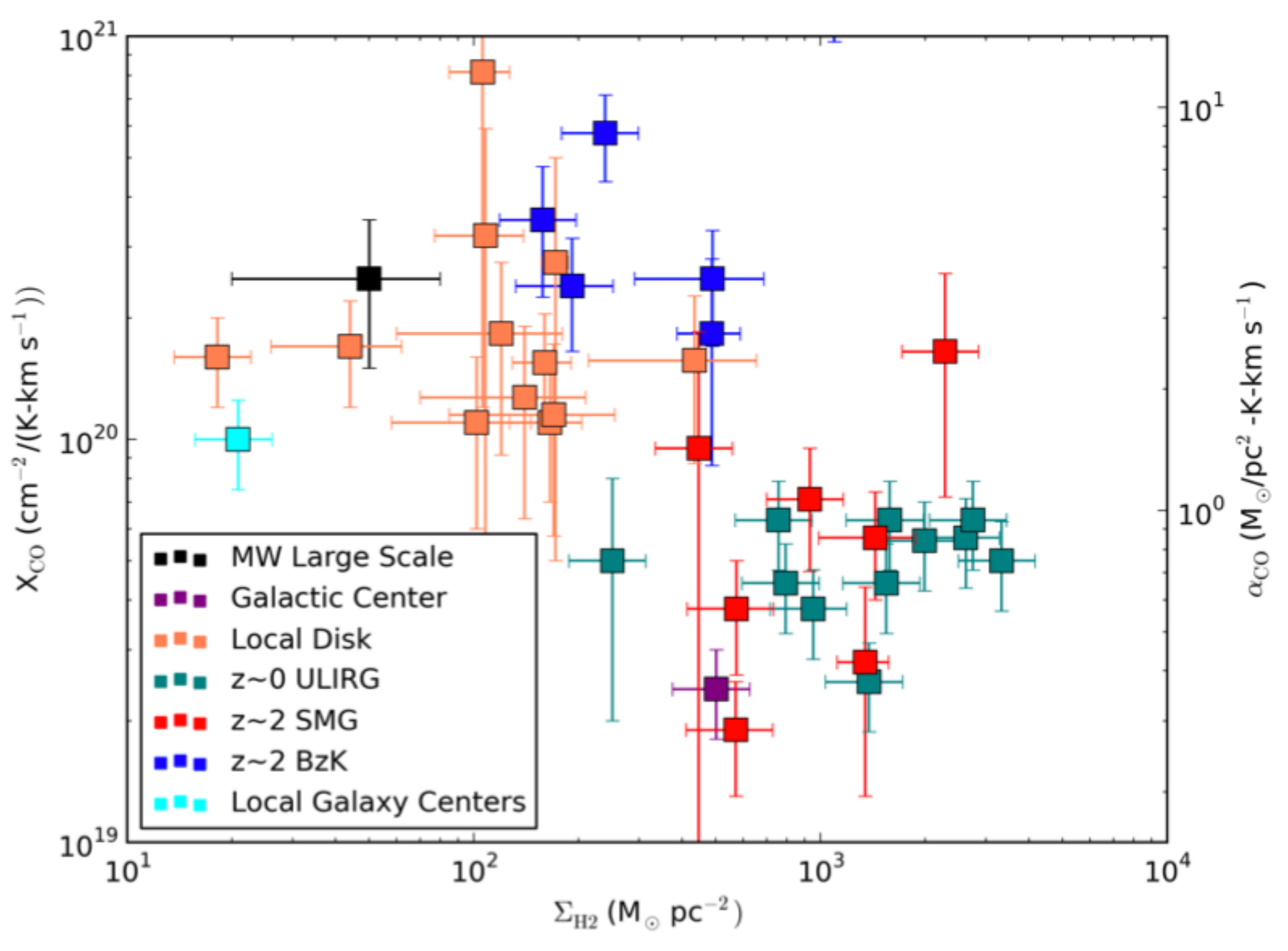}
\begin{spacing}{0.7}
\caption{\footnotesize An illustration of the diversity in CO spectral line energy 
distributions (SLEDs) for high-redshift galaxies (top) and existing
constraints on the X$_{\rm CO}$ (or $\alpha_{\rm CO}$) conversion
factor used to convert CO line luminosity to molecular hydrogen gas
mass (bottom).  The uncertainty in scaling from high-J CO lines to
CO(1-0) luminosity is a factor of $\sim$5 (from CO(3-2), and more at
higher-J transitions), and the uncertainty in $X_{\rm CO}$ is similar,
meaning gas masses derived from high-J transitions of CO alone will
necessarily be uncertain by factors of $\sim$25.  This margin of
uncertainty is 
astrophysically important quantities for understanding galaxy
evolution like gas fraction and total gas mass. 
the uncertainty is likely even larger for the lower mass galaxies
which ngVLA will readily study out to moderate redshifts,
and for which we have the most to learn about basic gas content
characteristics and interplay with the ISM.}
\end{spacing}
\label{fig:coexcite}
\end{figure}

The uncertainties involved in scaling high-J CO transitions to
molecular gas mass are illustrated in Figure~\ref{fig:coexcite}, which
shows both the diversity of CO spectral line energy distributions
(SLEDs) amongst high-redshift galaxies and the existing constraints on
the CO--H$_{2}$ conversion factor amongst high-redshift galaxy
populations.  When all uncertainties are accounted for, high-J
CO-derived gas masses are uncertain by factors of $\sim$25, an
uncertainty far too large to precisely measure important quantities
like galaxies' gas fractions.

Due to the uncertainties involved with converting high-J CO lines to
gas masses, recent work has focused on using dust continuum
measurements to scale to gas
masses \citep{magdis11a,magdis12a,scoville14a,santini14a}.  However,
this technique typically assumes that the dust-to-gas ratio is fixed across a
wide range of galaxies and redshifts, which has been shown to be a
poor assumption in some cases \citep{remy-ruyer14a,capak15a}.
In particular, the dust-to-gas ratio is known to depend strongly on metallicity \citep[e.g.,][]{issa90,Lisenfeld98,Draine07b}.
Furthermore, the important kinematic signatures that come along with
molecular line measurements are absent from dust continuum
measurements.

The ideal probe of the cold molecular gas reservoir is the ground
state transition of CO, CO(1-0), where the scaling to molecular
hydrogen mass has only the uncertainty in $X_{\rm CO}$ with which to
contend.  Furthermore, constraining $X_{\rm CO}$ more precisely in
high-$z$ galaxies is possible with constraints on dynamical mass and
stellar mass, which reduces the uncertainty on gas mass further.  The
JVLA, GBT, and ATCA have detected CO(1-0) in a number of high redshift
starburst galaxies, albeit a few at a time with long integrations
\citep{ivison11a,hodge12a,papadopoulos12a,greve14a,emonts14a}.  These
initial detections of CO(1-0) at high-redshift have not only
constrained the total molecular gas mass of hydrogen in these high-z
systems, but also highlighted the importance of understanding the
spatial distribution of gas in distant galaxies.

While the submillimeter-luminous galaxy population, i.e. dusty star
forming galaxies (DSFGs) have been found to be relatively compact in
high-J transitions \citep{tacconi08a,bothwell10a}, consistent with the
idea that they are scaled-up analogs to local
ULIRGs \citep{sanders96a}, their CO(1-0) maps are dramatically
different, showing gas extending as far as 16\,kpc from the galaxy
centers \citep{ivison11a,spilker15a}.  Studies of high-redshift
proto-cluster radio galaxies even revealed bright CO(1-0) reservoirs
out to distances of $\sim$60\,kpc from the central
galaxy \citep{emonts14a}.  While these larger sizes are more
suggestive of disk-like rotation dynamics and widespread molecular gas
reservoirs in the halo environment of massive galaxies, the number of
galaxies surveyed at the current VLA sensitivity makes it difficult to
infer large-scale population dynamics or draw comparisons between
low-$z$ and high-$z$ galaxy populations.

\begin{figure}[]
\vspace*{-1cm}
\centering
\includegraphics[width=0.8\textwidth,natwidth=612,natheight=792]{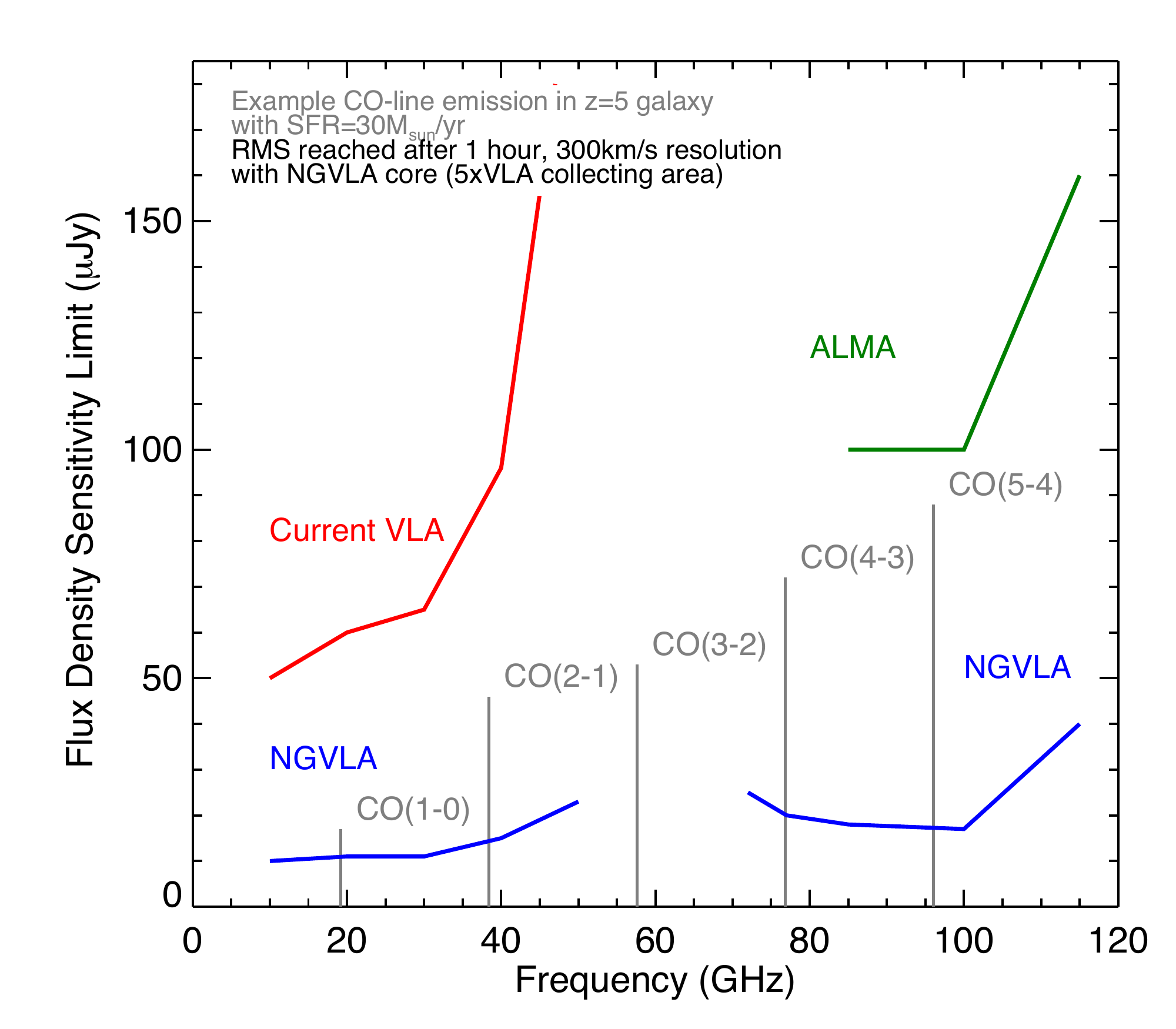}
\begin{spacing}{0.7}
\caption{\footnotesize The projected sensitivity limits of the proposed ngVLA core, 
which will have five times the effective collecting area of the
current VLA.  This plot also shows the sensitivity limits of ALMA
alongside the current VLA and projected ngVLA as a function of
frequency (the gap from 50--75\,GHz is due to atmosphere).
Overplotted are the anticipated CO line strengths, from CO(1-0) to
CO(5-4) of a typical $z=5$ galaxy (with SFR=30\,M$_\odot$\,yr$^{-1}$).
This type of galaxy is currently well below the detection limit of
existing facilities, including ALMA, yet is crucial for understanding
the star forming budget of the Universe's first galaxies.}
\end{spacing}
\label{fig:sensitivity}
\end{figure}

The ngVLA is desperately needed to dramatically increase the number of
galaxies surveyed in CO(1-0) at high-redshift $-$ by factors in the
thousands, in line with what is currently observed from high-redshift
galaxies via direct starlight.  Figure~\ref{fig:sensitivity}
illustrates the anticipated improvement in ngVLA sensitivity over the
current VLA and ALMA depths across 10--120\,GHz.

\subsection{A molecular gas deep field}\label{sec:deepfield}


\begin{figure}[]
\centering
\includegraphics[width=1.0\textwidth,natwidth=612,natheight=792]{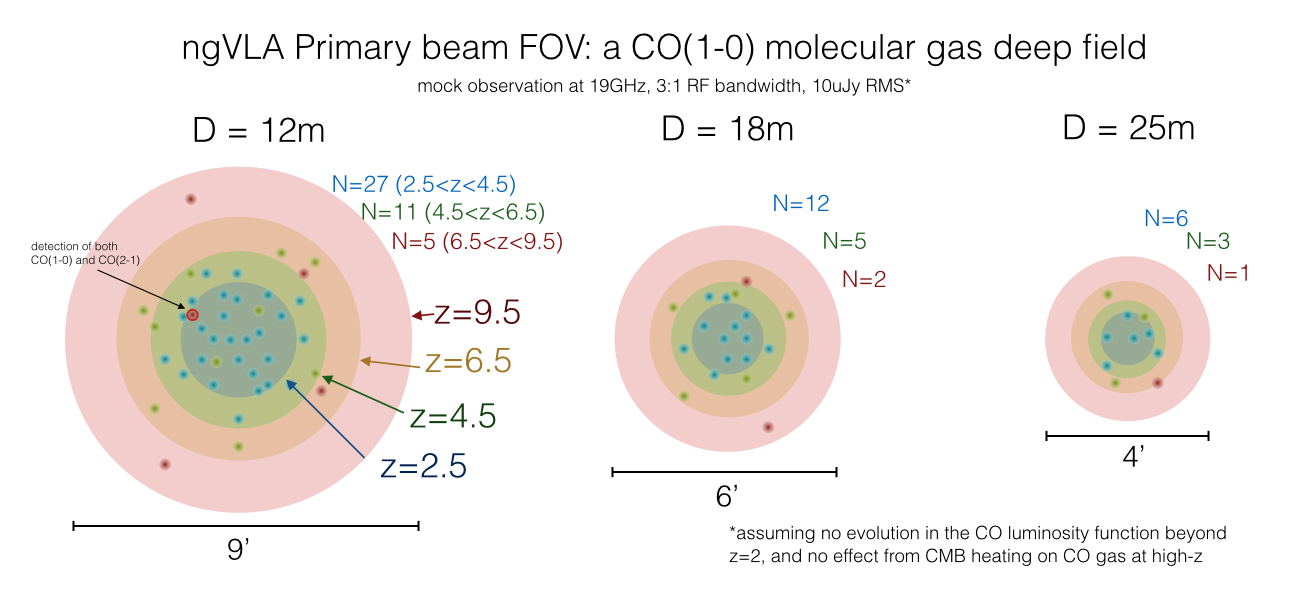}
\begin{spacing}{0.7}
\caption{\footnotesize A schematic of the field of view and spectroscopic line 
identifications for a single ngVLA pointing tuned to
$\nu$$=$11-33\,GHz (midpoint of log($\nu$) at 19\,GHz) with a total
3:1 RF bandwidth and 10\,\uJy\,beam$^{-1}$\ RMS.  The field of view is
primarily dependent on dish diameter. From the perspective of mapping
large regions of the Universe in CO, a dish size of 12\,m is clearly
favored.  The field of view also then depends on frequency and
redshift; here we denote the lowest frequencies with the pink circles
(corresponding to the frequency where we would detect CO(1-0) at
$z=9.5$), and higher frequencies in yellow (CO(1-0) at $z=6.5$), green
($z=4.5$), and blue ($z=2.5$).  Mock sources in three redshift bins
($2.5<z<4.5$ in blue, $4.5<z<6.5$ in green, and $6.5<z<9.5$ in red)
are overlaid.  Note that above $z>6.5$, this configuration is capable
of detecting both CO(1-0) and CO(2-1) if a source falls within the
central high-frequency field of view.}
\end{spacing}
\label{fig:deepfield1}
\end{figure}

To demonstrate the dramatic impact of ngVLA on detection rates of
CO(1-0) in high-$z$ samples, we have generated a mock observation with
a 3:1 bandwidth ratio tuned to 11--33\,GHz. This frequency range will
pick up the CO(1-0) transition from $2.5<z<9.5$. The RMS reached after
a one hour integration with five times the collecting area of the
current VLA is $\sim$10\,$\mu$Jy\,beam$^{-1}$ across the whole
bandwidth.  This means that in 30 hours (including overheads), ngVLA will be able
to achieve a 1$\sigma=$\,10\,$\mu$Jy\,beam$^{-1}$ RMS in 2\,MHz bins,
sufficient to resolve individual CO lines into 3--5 spectral bins
across the entire bandwidth.


To estimate the number of galaxies accessible to CO(1-0) detection
within one ngVLA pointing, we use the latest estimates of the
integrated infrared luminosity function \citep{casey14a} which is
roughly complete out to $z\sim2$.  For lack of a better assumption, we
assert that the same luminosity function holds out to $z\sim10$, and we 
translate $L_{\rm IR}$ to a best-guess CO(1-0) luminosity function
using a conservative estimate of a ULIRG-type L$_{\rm CO(1-0)}^\prime$-to-L$_{\rm IR}$ scaling \citep{bothwell13a} 
and assuming a linewidth of 200\,km\,s$^{-1}$.
Figure~\ref{fig:deepfield1} gives a schematic of detectable galaxies
within a single pointing to a fixed depth of 10\,\uJy\,beam$^{-1}$
RMS.  This schematic highlights two things: the field of view is
substantially larger for smaller-dish antennae (12\,m as opposed to
the current 25\,m), and the field of view is substantially larger at
low frequencies, corresponding to the regime where the
highest-redshift CO emitters would be detectable.  However, with a
fixed flux density RMS, only the brightest galaxies will be detectable
at the highest redshifts and so are comparatively fewer in number than
the fainter, $z<4.5$ sources.  For example, assuming 18\,m antennae,
we expect twelve $2.5<z<4.5$ sources to fall within a 2--3$'$ field of
view, five $4.5<z<6.5$ sources to fall within a 3--4$'$ field of view,
and two $6.5<z<9.5$ sources to fall within a 5--6$'$ field of view.

Recently, \citet{da-Cunha13b} published empirical predictions for the
number of line emitters in a CO deep field. Using the deepest
available optical/near-infrared data for the Hubble Ultra Deep Field
(UDF), they employed a physically motivated spectral energy
distribution model along with empirical relations between the observed
CO line and infrared luminosities of star-forming galaxies in order to
predict the expected number of line detections. In particular, for
frequencies between 18--50\,GHz over the 12 arcmin$^2$ UDF, and
assuming a linewidth of 300\,km\,s$^{-2}$, they predict of order
$\sim$200 CO(1-0) detections down to a 5$\sigma$ limit of
50\,$\mu$Jy\,beam$^{-1}$.  In order to compare the two methods, if we
adjust our frequency tuning to cover the same range, we predict
$\sim$70 CO(1-0) line detections in a single (12-m dish) pointing,
which is approximately one-third of the area of the UDF. The two
methods thus produce predictions that are of the same order. Given all
of the assumptions that went into the models, and particularly since
the estimates were derived in two completely different ways, this
relative agreement is encouraging.

The mock observation described in this paper 
will not only be sensitive to CO(1-0), but also
CO(2-1) emission at early epochs. However, the sensitivity and field
of view for CO(2-1) will be significantly different for the same
redshift regime given the differences in frequency, and will more
closely mirror the coverage for the low-redshift CO(1-0) emitters.
The survey sensitivity with respect to CO(2-1) depends on source
excitation as the ratio of $S_{\rm CO(2-1)}/S_{\rm CO(1-0)}$ can vary
from 2--4.  Perhaps counter-intuitively, the depth of CO(2-1)
observations at high-$z$ will be greater than for CO(1-0), although
the number of expected sources per single pointing will not differ
substantially from the expected CO(1-0) emitters given the limited
field of view.  Line degeneracy may then be of concern, since it will
be difficult to disentangle low-redshift CO(1-0) emitters and
high-redshift CO(2-1) emitters.  Other multiwavelength information
(and perhaps parallel ALMA observations) would be necessary to break
this degeneracy.

\begin{figure}[]
\centering
\includegraphics[width=0.7\textwidth,natwidth=612,natheight=792]{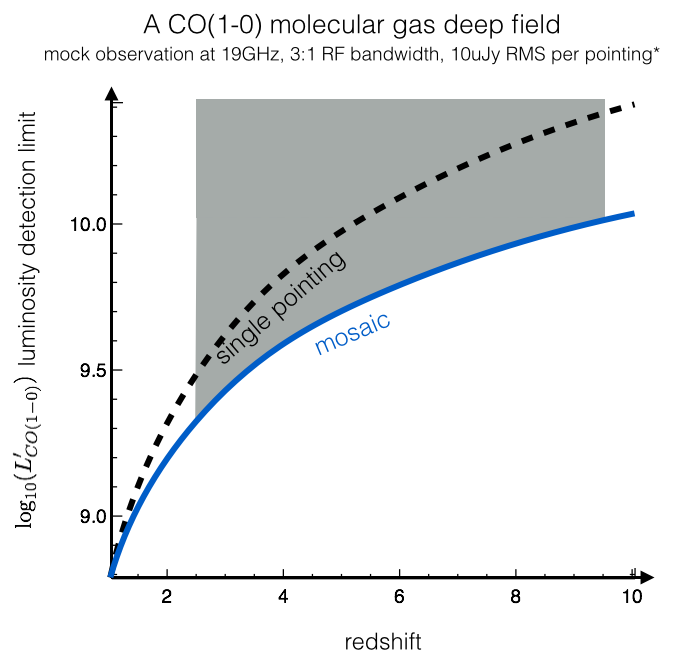}
\begin{spacing}{0.7}
\caption{\footnotesize The depth of a fixed-10\,\uJy\,beam$^{-1}$ 
RMS CO(1-0) search as a function of redshift for a single pointing
(dashed line) vs. a mosaic (blue).  The units on the y-axis
$L^\prime_{\rm CO(1-0)}$ are K\,km\,s$^{-1}$\,pc$^2$; assuming a
median value of $\alpha_{\rm CO}\approx2$\,M$_\odot$/L$^\prime_{\rm
CO(1-0)}$, a y-axis value of 9.0 corresponds to a gas mass of roughly
2$\times$10$^9$\,M$_\odot$, and 10.0 is a gas mass of roughly
2$\times$10$^{10}$\,M$_\odot$.  The redshift range probed by CO(1-0)
at these frequencies is indicated by the gray region.  Using the high
end of the frequency range to define the spatial distance between
pointings causes significant spatial overlap for lower frequencies,
pushing the RMS in those areas much lower and thus pushing the
high-redshift discovery space deeper.}
\end{spacing}
\label{fig:deepfield3}
\end{figure}

Note that weaving a series of single pointings together will have
unique advantages with such a wide bandwidth.  For a single pointing
example above, the low frequency space probes the largest volumes, but
at shallower luminosity limits at high-redshifts.  Tiling a mosaic
together by maximizing overlap for the low-frequency observations will
dramatically reduce the RMS at higher-redshifts, while mapping a large
area.  Figure~\ref{fig:deepfield3} shows the effect of mosaicing on
the CO(1-0) luminosity limit out to high-redshifts, sensitive to
objects in the gray parameter space.  As an example, we predict that a
$\approx$750\,hour mosaic made up of 41 individual pointings with
12\,m antennae (5$\times$5 pointings overlapped with an inset
4$\times$4 pattern) could detect nearly 1500 individual CO(1-0)
emitters, with the largest number above $z>5$ where the mosaic would
be at its deepest.  This mosaic would be approximately
20$'\times$20$'$ across.  In contrast, antennae which are 18\,m in
size would yield only $\sim$600 detections, and 25\,m dishes would
yield only $\sim$300 detections.


%

While direct molecular gas detections are certainly the main focus of
future ngVLA extragalactic deep fields, it is worth noting that these
surveys will be done in well-surveyed legacy fields, where there is
already substantial ancillary data including spectroscopic redshifts.
With large numbers of spectroscopic redshifts, stacking to measure the
median molecular gas reservoir in LBGs would become possible as a
function of redshift, environment, or other physical factors.

Before leaving this topic, it is important to return to our {\it a
priori} assumption that the luminosity function of CO will not
substantially change between $z=2.5$ and $z\sim10$.  This undoubtedly
is not the case, but as there are currently no constraints on any alternative
picture, we cannot hypothesize on the 
change we will perceive in
detection yields in molecular gas deep fields.  In addition, it is
likely the case that early Universe galaxies contain fewer metals and
may be less luminous in CO for their given mass.  A final important
caveat is that at sufficiently high redshift, the temperature of the
CMB is non-negligible and can contribute to dust continuum and gas
heating within galaxies. 
While this extra heating of the gas (and dust) will boost the intrinsic fluxes, the CMB also becomes a stronger background against which these fluxes must be measured. 
The net effect of these two competing effects on CO emission is
less straightforward than for the dust continuum emission \citep{dacunha13a}, as the
regions of cold molecular gas may not be in local thermal equilibrium,
or the kinetic temperature of the gas may not be thermally coupled to
the dust in an obvious way, particularly in early-Universe galaxies.
However, the overall effect of CMB heating will be to cut down the number of direct
detections in the $z\simgt6$ Universe at low gas temperatures
($T\simlt20$\,K).

On a technical note, using phased array feeds (PAFs) with multiple
pixels instead of single pixel feeds would allow the ngVLA to survey
areas of sky corresponding to volumes of $\sim$ 0.1 Gpc$^3$, covering
the full range of cosmic environments. For example, if we were to use
a 3$\times$3 pixel PAF we could survey the entire 2\,deg$^2$ COSMOS
field to this depth in a few hundred hours, detecting $\sim 10^5$ CO
emitting galaxies. This would allow us, for example, to compute 2-point
correlation functions within the CO emitting population, and
cross-correlations between CO emitting galaxies and other populations
such as galaxies with large stellar masses, galaxy clusters and AGN,
all of which can be compared directly with the output of simulations.
This is critical for assessing the collapse of large scale structure
and environmental impact on star formation in galaxies.

\subsection{Using CO as a redshift beacon}


The need for CO-based redshift confirmation for the most dust-obscured
starburst galaxies has already been made clear by the past decade of
effort poured into obtaining spectroscopic confirmation for dusty
star-forming
galaxies \citep[DSFGs; e.g.,][]{chapman05a, swinbank04a, casey11a, Casey12a, Casey12b, danielson15a}.
These studies have clearly demonstrated that $\sim$50\%\ of all dusty
starbursts are too obscured to obtain spectroscopic confirmation via
emission lines in the optical/near-infrared.  The question of whether
the unconfirmed DSFGs sit at similar redshifts or higher redshifts
than the others is yet to be clarified, but it is quite clear that
DSFG samples beyond $z>3$ suffer from severe spectroscopic
incompleteness.  The discovery of bright dusty galaxies like
HFLS3 \citep{riechers13a} hint at the existence of such a population
at very high redshifts, but so far, confirming such sources (at more
typical, unlensed luminosities) in a systematic way through blind CO
searches has not been efficient.  In fact, even the existing
capability of ALMA can only complete a spectral scan for high-J CO emission for
a single source in about an hour of integration time -- still prohibitive for
large samples, save the strongly lensed subset \citep{weis13a}.
If we consider the mock observation in \S~\ref{sec:deepfield} and
Figure~\ref{fig:deepfield1}, we expect to detect $\sim$100 galaxies in one
pointing with a 3:1 RF bandwidth ratio above a CO line luminosity of
$L_{\rm CO(1-0)}^\prime=10^{9.2}$\,K\,km/s\,pc$^2$.  Those CO
detections can potentially probe a redshift range as wide as
$2.8<z<10.5$.  This makes the case clear that NGVLA will be
revolutionary in providing unequivocal spectroscopic confirmation for
galaxies containing substantial molecular gas potential wells,
potentially surpassing the optical/near-infrared in efficiency of
follow-up.

\begin{figure}
\centering
\includegraphics[width=0.49\textwidth,natwidth=612,natheight=612]{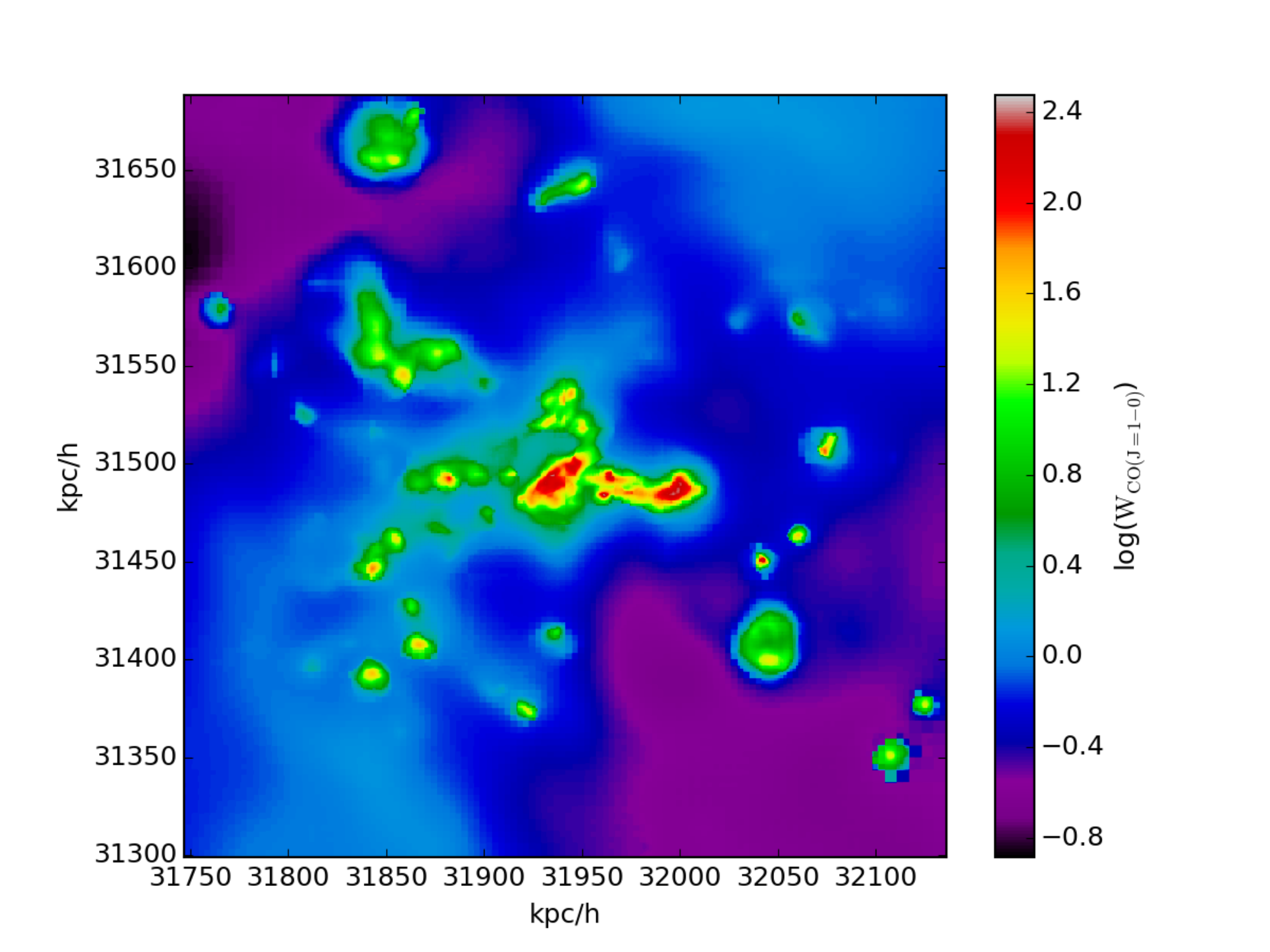}
\includegraphics[width=0.49\textwidth,natwidth=612,natheight=612]{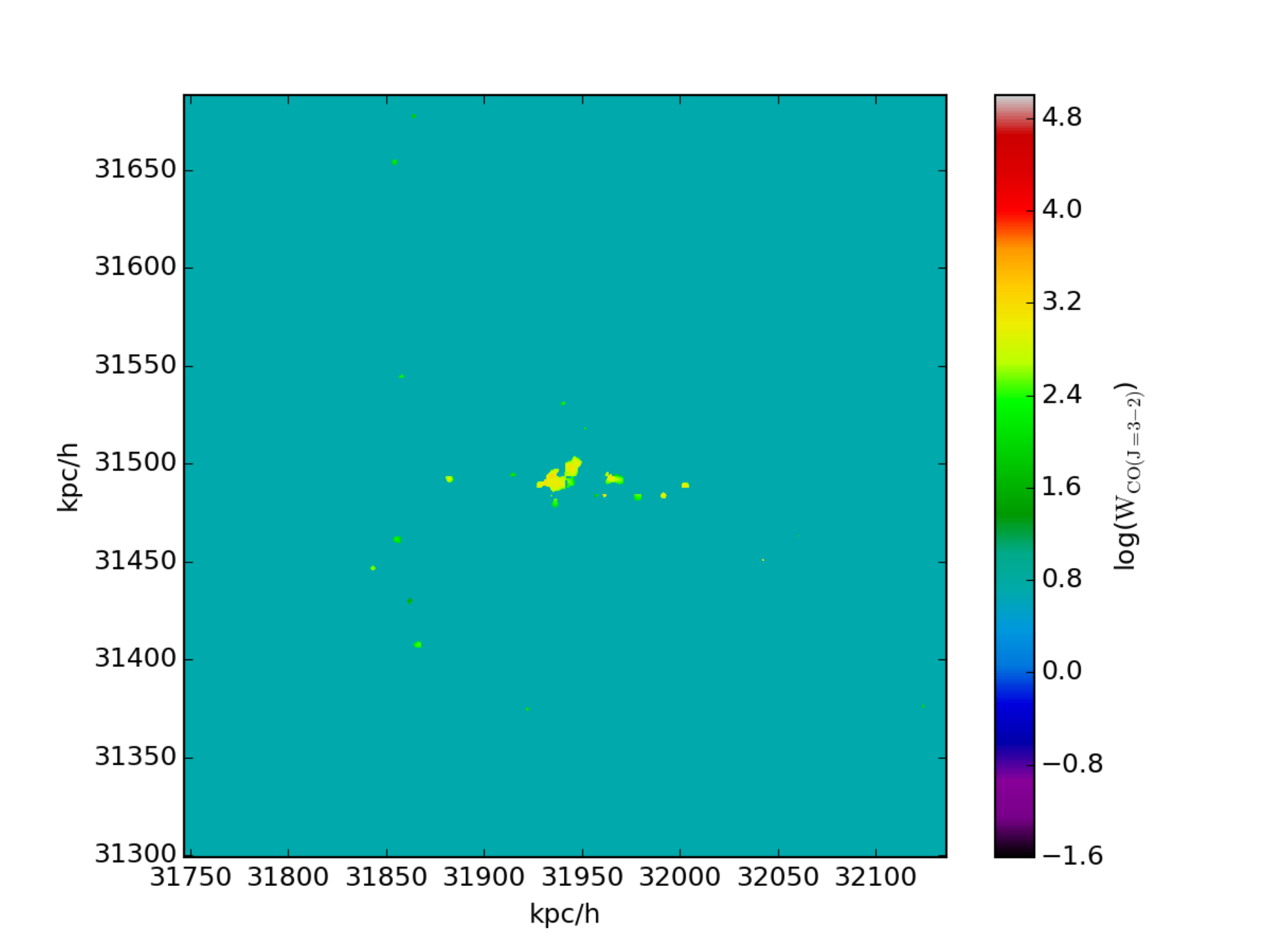}
\begin{spacing}{0.7}
\caption{\footnotesize Simulated CO(1-0) emission (left) and CO(3-2) 
emission (right) from a model z$\sim$2 dusty star forming galaxy in
  high-resolution cosmological zoom
  simulations \citep{narayanan15b}. The units of W$_{CO}$ are
  K\,km\,s$^{-1}$. Dusty star forming galaxies trace rich environments
  with central galaxies being bombarded by subhalos and diffuse gas.
  Because much of the gas in the large scale environment is below the
  effective excitation conditions for dense gas tracers (even
  CO(3-2)), probing the environment of the most extreme star-formers
  in the Universe at high-redshift will require observing the ground
  state CO transition, with a compact configutation sensitive to
  large scale structures.  }
\end{spacing}
\label{fig:codn}
\end{figure}

\begin{figure}
\centering
\includegraphics[width=1\textwidth,natwidth=612,natheight=612]{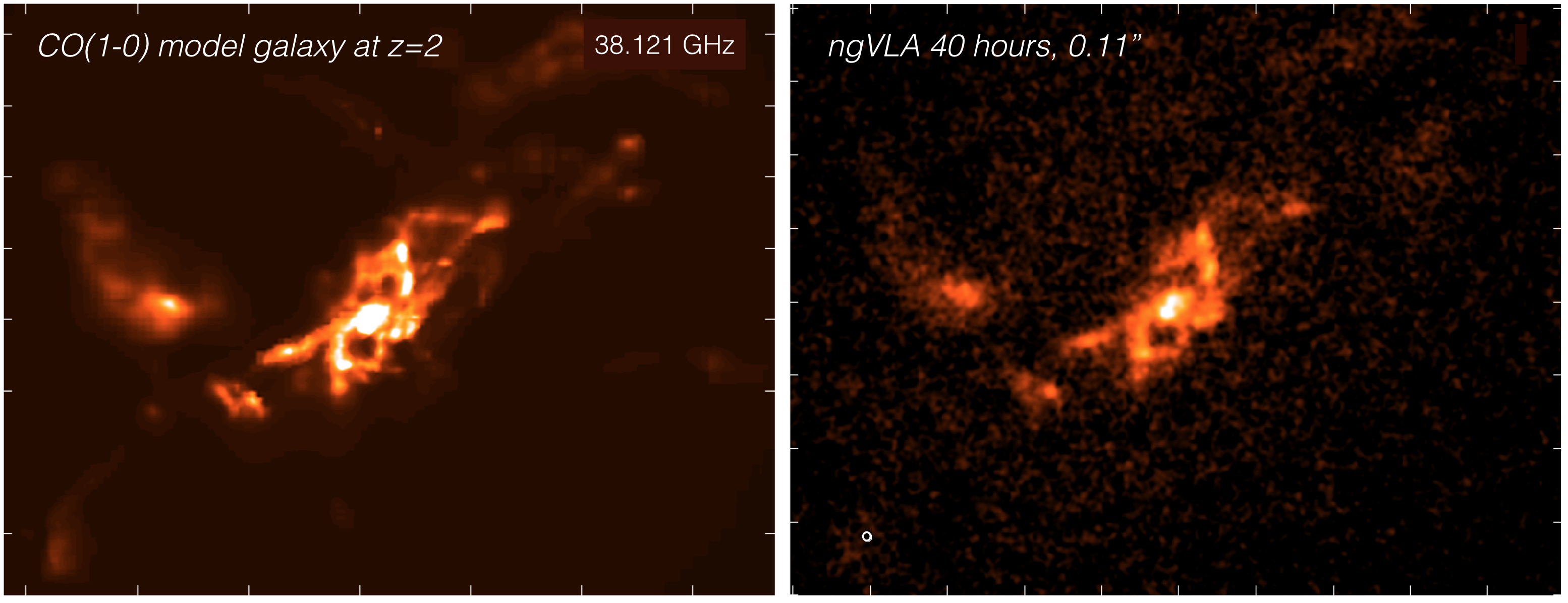}
\begin{spacing}{0.7}
\caption{\footnotesize Illustration of ngVLA capabilities to detect 
CO(1-0) emission from a idealized model galaxy at $z=2$.  In 40 hours
of integration at 0.11$''$ resolution, ngVLA will be able to resolve
great structural detail in the molecular gas reservoir of high-$z$
systems, which is currently inaccessible to both ALMA and the current
JVLA.  In this simulation, the ngVLA reaches an RMS gas mass
sensitivity of 3$\times$10$^8$\,$(\alpha_{\rm CO}/4)$\,M$_\odot$ at
1\,kpc resolution; this is 10 times deeper than the current VLA would
reach in the same time.
}
\end{spacing}
\label{fig:codncc}
\end{figure}

\subsection{Simulations of CO gas emission at high-$z$}

Dusty star-forming galaxies at high-redshift are the most luminous,
heavily star-forming galaxies in the Universe \citep*{casey14a}.  As
some of the most luminous star-formers, ngVLA is in a unique position
to study DSFGs' CO emission in detail through high-resolution, pointed
follow-up.

Theorists argue at great length over physical models for DSFGs, with
diverse predictions existing for their masses, large scale
environments, and potential merger status.  Broadly, models for the
origin of DFSGs fall into two camps: one in which major mergers drive
the luminosity of the galaxy, and one in which DSFGs are the centers
of massive halos growing hierarchically via numerous minor mergers and
smooth accretion.

High-resolution observations may have the power to distinguish between
these two broad classes of models.  For example, high resolution
cosmological zoom simulations \citep{dave10a,narayanan15a} suggest
that DSFGs owe their origin to hierarchical growth in a LCDM cosmology
will show significant substructure around the central galaxy when
observed at high-resolution.  However, much of this gas associated
with subhalos may be relatively diffuse compared to the relatively
dense gas in the central halo.  Figure~\ref{fig:codn} presents
synthetic CO emission from one such model for DSFG formation
\citep{narayanan15b}.  The subhalos being accreted onto the central
halo are clearly present in the ground state CO(1-0) transition.
However, even CO(3-2) has an effective density too large to light up
in the diffuse subhalos.

Because of this, (sub)mm-wave interferometers such as ALMA may miss
the large scale gaseous substructure in high-z DSFGs, necessitating
the use of sensitive radio-wave interferometers like the ngVLA. In
this respect, recent observational work on low-surface-brightness
CO(1-0) emission in and around high-z proto-cluster radio
galaxies \citep{emonts14a,emonts15a}, already revealed very widespread
($\sim 60$\,kpc) reservoirs of diffuse molecular halo gas, reminiscent
of the simulations shown in Figure~\ref{fig:codn}. These
low-resolution observations were done with the Australia Telescope
Compact Array, with its five antennas packed in ultra-compact array
configurations that surpass those of the current VLA. The ngVLA can
revolutionise the studies of low-surface-brightness CO(1-0) emission
with a densely packed core that consists of many short-baseline
elements, with more than an order of magnitude more sensitivity than
existing compact arrays.

Figure~\ref{fig:codncc} shows this same hydrodynamically simulated
$z=2$ galaxy as shown in Figure~\ref{fig:codn}, but with simulated
noise properties characteristic of the ngVLA 40 hours of integration
at 0.11$''$ resolution.  Such observations of high-$z$ systems is
currently quite limited with the JVLA (GN20, shown in
Figure~\ref{fig:gn20}, is exceptionally luminous and required
120\,hours of observations to reach adequate sensitivity).  The
sensitivity improvements with ngVLA will not only cut the time per
source substantially but reach much more intrinsically faint systems.

\subsection{Dense gas tracers}\label{sec:densegas}  

The critical density required to collisionally excite CO is relatively
low (n$_{\rm H_2}$$\sim$10$^{2}$--10$^{3}$ for the lower-J
transitions), meaning that observations of CO are a good way to trace
the total molecular gas reservoirs in galaxies. At the same time, this
also makes CO a fairly poor tracer of the dense molecular cores where
star-formation within distant galaxies is ultimately taking
place. High dipole moment molecules like hydrogen cyanide (HCN), on
the other hand, are only collisionally excited at very high densities,
making them much more reliable tracers of the very dense gas directly
associated with the formation of individual stars. Some studies of HCN
in the nearby universe have even found evidence that the ratio of HCN
luminosity to FIR luminosity remains constant over $>$8 orders of
magnitude in HCN luminosity, suggesting that HCN may be a fundamental
direct probe of star forming `units,' and that the only difference
between star formation on different scales and in different
environments is the number of these fundamental star-forming
units \citep[e.g.][]{gao04a,wu05a,wu10a,zhang14a}.

\begin{figure}
\centering
\hspace{-1cm}\includegraphics[width=1.0\textwidth,natwidth=800]{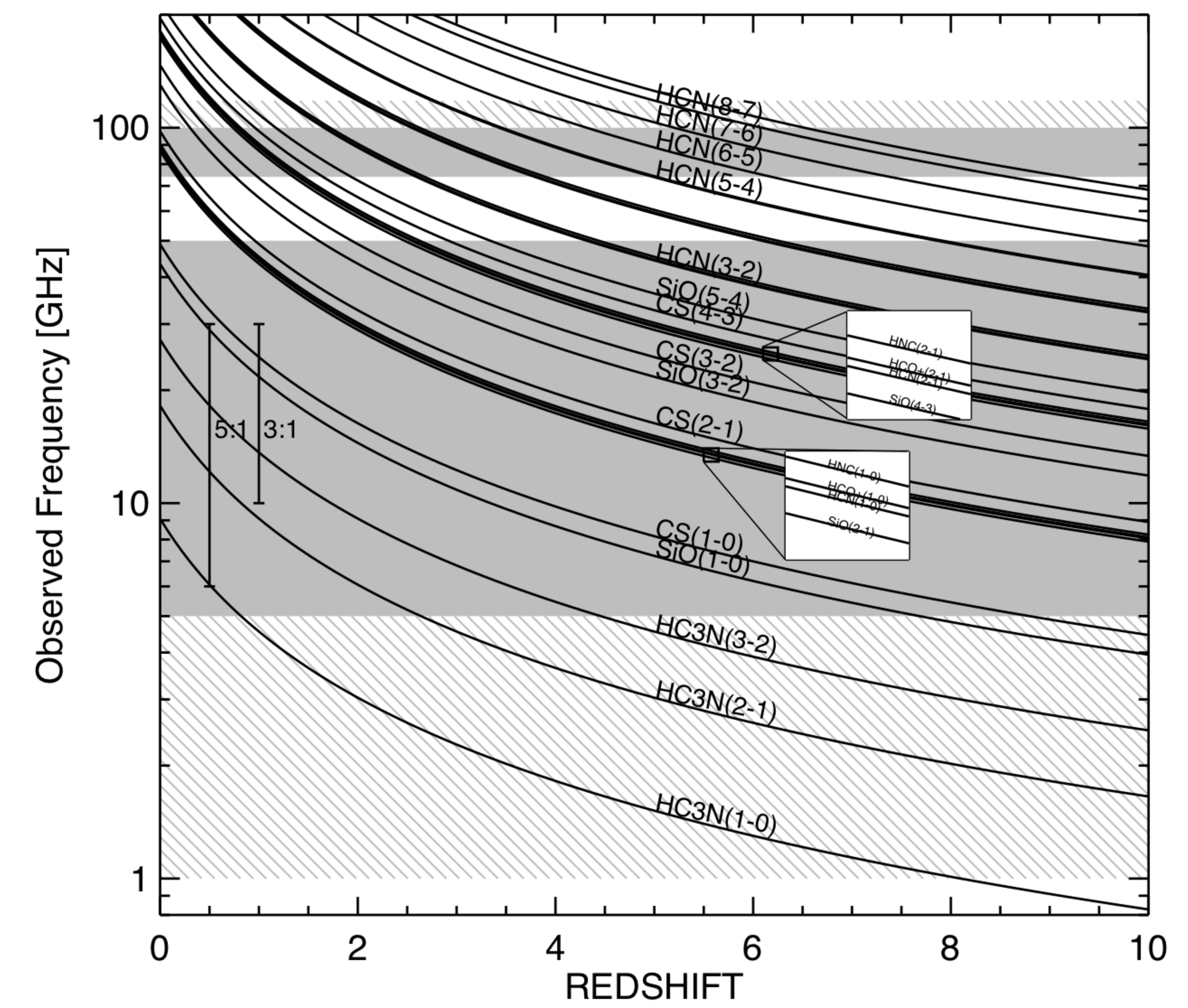}
\begin{spacing}{0.7}
\caption{\footnotesize Redshifted frequencies of dense gas tracers,
  many of which will be accessible in single-tuning setups with the
  NGVLA at high-redshift. In particular, the 1-0 and 2-1 transitions
  of HNC, HCO$+$, HCN and SiO lines are spaced very closely in
  frequency (see inset zoom-ins).  While ALMA will be able to detect
  the mid- and high-J transitions of these molecules at high-redshift,
  significant uncertainties regarding excitation
%
%
mean that 
the low-J transitions accessible with the ngVLA more suitable as tracers of the total dense gas mass. 
%
%
}
\end{spacing}
\label{fig:densegas}
\end{figure}

While some recent observational studies argue that the local HCN-FIR
relation may in fact be more
nuanced \citep[e.g.][]{garcia-burillo12a} -- a possibility for which further investigation by the ngVLA will be essential -- 
it is clear that even less
is known about dense gas tracers in the more distant universe.  Owing
to the fact that they only trace the densest regions of the ISM and
are therefore less abundant, emission from the rotational transitions
of molecules like HCN is usually an order of magnitude fainter than
CO, complicating efforts to detect and study these tracers at
high-$z$. As a result, only a few high-z galaxies have been detected
in dense gas tracers to-date 
\citep[e.g.][]{solomon03a,vanden-bout04a,carilli05a,gao07a,riechers07a,riechers11a,danielson11a}.
With ALMA now online, the situation will clearly improve dramatically
in the near future. However, as with CO, ALMA will only be able to
detect the mid- and higher-J transitions of dense gas tracers like
HCN, HNC, and HCO+. These higher-level transitions are less directly
tied to the total dense gas mass, requiring assumptions about the
(highly uncertain) excitation ratios. In addition, these higher-J
transitions are more likely to be affected by IR pumping, which local
studies find may be common in ultraluminous
galaxies \citep[e.g.][]{aalto07a}.  Thus, while necessary to
understand the overall excitation properties of high-$z$ sources,
the ALMA-detectable transitions may be unsuitable as tracers of the 
total dense gas mass.

The current JVLA probes the right frequency range to detect the
crucial low-J transitions of these high critical density molecules
(Figure~\ref{fig:densegas}). However, its limited sensitivity means
that the current state-of-the-art for high-z detections consists of a
smattering of strongly lensed hyper-starbursting quasar
hosts \citep[e.g.][]{vanden-bout04a,riechers07a,riechers11a}.  With
significantly increased sensitivity, the ngVLA would extend studies of
the dense gas mass at high-z beyond this handful of extreme objects
for the first time. In addition to tracing the dense gas mass at
high-z, such studies are critical for constraining models of star
formation based on the gas density PDF, as these models make testable
predictions about the FIR-HCN relation in FIR-luminous
objects \citep[e.g.][]{krumholz07a,narayanan08a}. Finally, while
angular resolution is not a priority for these photon-starved studies,
the brightest objects could even be spatially resolved by a ngVLA on
$\sim$kpc scales (requiring baselines on the order of the current
VLA). The ngVLA would thus enable detailed studies of the dense gas at
high-redshift such as are currently only possible in the local
universe.

\section{Dynamics}


\label{sec:dynamics}

Of the several hundred $z>1$ galaxies currently detected in molecular
line emission \citep[see][for a review]{carilli13a}, the vast majority
are spatially unresolved or -- at best -- marginally
resolved. Particularly in the low-J (J$<$3) CO transitions which trace
the bulk of a galaxy's gas reservoir, only a handful of high-redshift
galaxies have been resolved
to-date \citep{riechers08a,hodge12a,hodge13a}. This lack of spatial
resolution makes it very challenging to do any detailed dynamical
modeling.  By providing sub-arcsecond resolution imaging of the gas
and dust in high-redshift galaxies, the ngVLA will allow detailed
studies of morphologies and -- even more critically -- galactic
dynamics. This section discusses some of the applications this ability
will have for studies of galaxy evolution at high-redshift.

\subsection{The relative role of mergers versus disks}

\begin{figure}
\centering
\includegraphics[natwidth=600,width=1.0\textwidth]{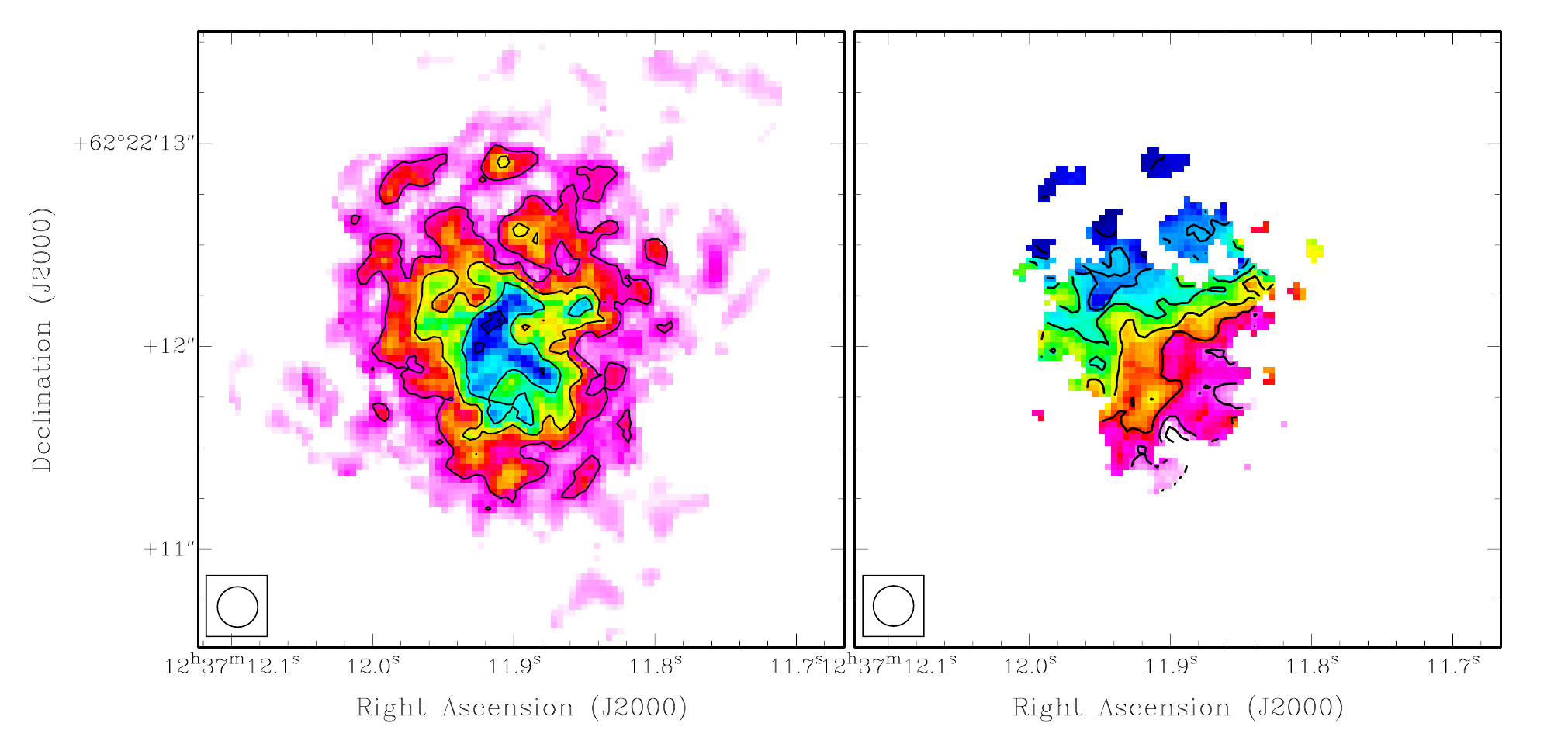}
\begin{spacing}{0.7}
\caption{\footnotesize
High-resolution VLA CO(2--1) moment maps for the $z=4.05$
submillimeter galaxy (SMG) GN20 \citep{hodge12a}. Using the VLA's B$+$D
configurations, \citeauthor{hodge12a} achieved a resolution of
$\sim$1.3 kpc, allowing them to resolve the extended cold gas
reservoir as well as model its dynamics. This is the most detailed map
of a high--z SMG to-date, but it required 120 hours of VLA time (50
hours on-source) on this extremely gas-rich source. The NGVLA is
critical to extend such studies beyond a handful of extreme
high-redshift sources.}
\end{spacing}
\label{fig:gn20}
\end{figure}

What is the relative role of major mergers versus secular disk
evolution to galaxies in the early Universe? This question applies not
only to the most massive, dust-obscured
starbursts \citep[e.g.][]{narayanan10a,dave10a,hayward12a}, but also
to the general mass assembly of galaxies at
$z>1$ \citep[e.g.][]{forster-schreiber09a,dekel09a}, when half of the
stellar mass in galaxies was assembled.  While the JVLA can already
spatially resolve emission on scales as small as 0.05$''$ (equivalent
to $\sim$400\,pc at a redshift of $z\sim1$), it lacks the surface
brightness sensitivity to use this capability on the faint, extended
gas reservoirs in even the most CO-luminous high-redshift
galaxies. The best examples to-date of resolved CO at
high-redshift \citep[e.g.][]{riechers08a,hodge12a} therefore settle
for the VLA's B-configuration (rather than the more extended
A-configuration), reaching angular resolutions of $\sim$0.2$''$. This
resolution allowed \citet{hodge12a} to model the dynamics of the
$z=4.05$ DSFG named GN20, finding evidence for a rotating disk with a
flat rotation curve (see Figure~\ref{fig:gn20}). However, even in this
more compact configuration, the observations of this extremely
gas-rich galaxy (M$_{H_2}$ $>$ 10$^{11}$ M$_{\odot}$) required 50
hours on source, or a total of $>$100 hours with overheads. It is thus
currently not possible to do this experiment on more than a handful of
the most extreme high-redshift galaxies (this is also demonstrated by
the example simulation in the previous section, shown in
Figure~\ref{fig:codncc}). Thus, even though the angular resolution of
the JVLA is already adequate for studies of high-$z$ galactic dynamics,
an increase in the surface brightness sensitivity by an order of
magnitude would be required for the ngVLA to feasibly conduct such
studies.

The newly completed ALMA has both the resolution and sensitivity
required to study the resolved morphologies and dynamics of some
high-redshift galaxies. However, as discussed in Section~\ref{sec:co},
ALMA is only able to detect the high-J transitions of CO in
high-redshift galaxies (see Figure~\ref{fig:coladder}), with CO(3-2)
and CO(5-4) the lowest transitions accessible at $z=3$ and $z=6$,
respectively (assuming the current bands). Beyond the well-known
uncertainties in global excitation ratios (Figure~\ref{fig:coexcite}),
the higher-J (J$\ge$3) lines have been shown to give a biased and
incomplete view of the reservoirs in some high-redshift galaxies, as
shown previously in Figure~\ref{fig:codn}. In particular, observations
of the CO in normal star-forming galaxies and DSFGs demonstrate that
CO(3--2) emission can miss up to \textit{half} of the gas mass and
imply reservoir sizes that are \textit{three times} more compact than
determined from the ground-state CO(1--0)
transition \citep[e.g.][]{hainline06a,ivison11a,dannerbauer09a,aravena10a,riechers11a}. It
is thus crucial to extend down to the frequency range of the VLA
(i.e., down to $\sim$10-20GHz) in order to probe the dynamics of the
bulk of the gas directly.

Note that the [{\sc Cii}] emission line at a rest-frame 158\,$\mu$m
has been quite successful in characterizing the dynamics of some
high-redshift galaxies, although the interpretation is difficult as
the photo-disassociation region (PDR)-generated line's emission
represents a complex mix of physical mechanisms \citep{olsen15a}.
While the kinematics of the [{\sc Cii}] line is certainly a useful
diagnostic of potential mergers, it is unlikely to trace the full
extent of a galaxy's cold gas reservoir.  In contrast, the CO(1-0)
line provides not only dynamics, but a direct link to the molecular
gas supply, as detailed in the previous section.

\subsection{Molecular outflows}

In addition to distinguishing between mergers and disks, the dynamical
information can shed light on the presence of galactic
inflows/outflows. Galactic winds, in particular, are thought to play a
major role in regulating the formation of stars by acting as a source
of negative feedback on the star-forming ISM. As such, they are now a
common feature in theoretical models of galaxy evolution. By observing
the molecular component of such winds, one can measure the total
molecular mass outflow rate, which, along with the cool atomic phase,
is thought to comprise a significant fraction of the
wind \citep[e.g.][]{walter02a,rupke05a,alatalo11a}. Recent studies
with ALMA have mapped galactic winds in the local Universe in
unprecedented detail \citep[e.g.][]{bolatto13a} and, by detecting the
redshifted emission from low-J CO transitions in distant galaxies, the
ngVLA would allow studies of the importance of galactic outflows
across cosmic time.

\subsection{Measuring the CO-to-H$_2$ conversion factor}

Finally, the dynamical information is also critical to get back to the
physical parameter that is ultimately of concern: the total molecular
gas mass. Even with the ground-state transition of CO, an assumption
is required about the CO--to--H$_{\rm 2}$ conversion factor X$_{\rm
CO}$ in order to convert the observed CO luminosity to the total
molecular gas mass, M$_{\rm H_2}$. Locally, much progress is being
made in directly measuring the conversion factor through various
methods, including applying virial techniques to resolved CO emission
from discrete molecular clouds \citep[][ and references
therein]{bolatto13a}. Unfortunately, the direct determination of the
conversion factor in high-$z$ objects remains extremely challenging
with current instruments. Thus, the best we can do at high-$z$ right
now is to extrapolate from what we know about local galaxies, which
may not be appropriate analogues for the galaxy populations that exist
at earlier cosmic times. If our community is able to build the ngVLA,
that could achieve resolutions of $\sim$5-15mas in the (unbiased)
low-J CO lines, we could begin resolving individual molecular clouds
on scales of $\sim$40-100 pc at $z\sim4$, allowing us to measure the
conversion factor in high-$z$ galaxies directly.

\section{Continuum Emission}


Single-dish submillimeter imaging has contributed immeasurably to the
study of high-redshift galaxy evolution, particularly by unveiling a
significant population of extremely dusty star-forming galaxies that
have been incredibly challenging for cosmological simulations to
explain \citep[and references therein]{casey14a}.  This submillimeter imaging can cover large fields and
produce uniformly selected samples of distant DSFGs chosen in a single
bandpass to a well-established flux limit, making the selection of
these sources straightforward to model.  These wavelengths are also
equally sensitive to both low and high redshift galaxies, due to the
very negative $k$-correction on the Rayleigh-Jeans tail of dust
emission's blackbody.  However, images from single-dish submillimeter
facilities have low resolution, and blank field observations run up
against the confusion limit quickly.  Interferometric
submillimeter/millimeter imaging has high spatial resolution and
sensitivity, but the field-of-view is small. Radio continuum imaging
surveys with the VLA do not suffer from extinction, have high spatial
resolution, and can cover large areas.  However, at the present time,
the deepest surveys made with the VLA can only find the most luminous galaxies at
high redshifts, with star-formation rates above 100\,\msun\,yr$^{-1}$.
Here we discuss various ways in which continuum work with the ngVLA
will push the boundaries of galaxy evolution studies by probing the
dust-unbiased star formation from galaxies at high redshift.

\subsection{Synchrotron as a tracer of cosmic star formation}

Unlike the rest-frame UV/optical, synchrotron emission is relatively
unaffected by dust attenuation, making it a valuable tracer of the
total amount of star formation in galaxies, unbiased by dust.  The
main downside to using synchrotron emission as a star formation tracer
is the limiting sensitivity of current facilities.  For example, the
deepest existing 1.4~GHz VLA image ($5\sigma$ of 11.5\,$\mu$Jy of the
CDF-N by F. Owen) only probes down to ULIRG luminosities
(ultra-luminous infrared galaxies, L$_{\rm IR}=$\,10$^{12}$\,\lsun) to
$z\sim3$ and LIRG luminosities (luminous infrared galaxies,
10$^{11}$\,\lsun) to $z\sim1$.  The ngVLA can help us construct a
complete star formation history by delivering a large homogeneous
sample that is insensitive to extinction over a wide range of
luminosities and SFRs out to high redshifts (e.g., these same
luminosity limits to $z\sim5-6$).  With its unprecedented sensitivity,
the ngVLA will even be sensitive to nearly Milky Way-like galaxies at
redshifts of z$\sim$1--3, the peak epoch of galaxy formation.  For
example, using a 5--15\,GHz band (3:1 instantaneous RF bandwidth) and assuming a sensitivity
5$\times$VLA (10$\times$VLA) would result in a 5$\sigma$ detection
limit of 500\,nJy beam$^{-1}$ (250\,nJy beam$^{-1}$) after only 10
hours of on-source integration time.  Assuming baselines where the
source is unresolved, this translates to a SFR of
10.8\,M$_{\odot}$\,yr$^{-1}$ (5.4\,M$_{\odot}$\,yr$^{-1}$) at
$z\sim2.5$.

\begin{figure}
\begin{center}
\includegraphics[natwidth=4in,width=4.0in]{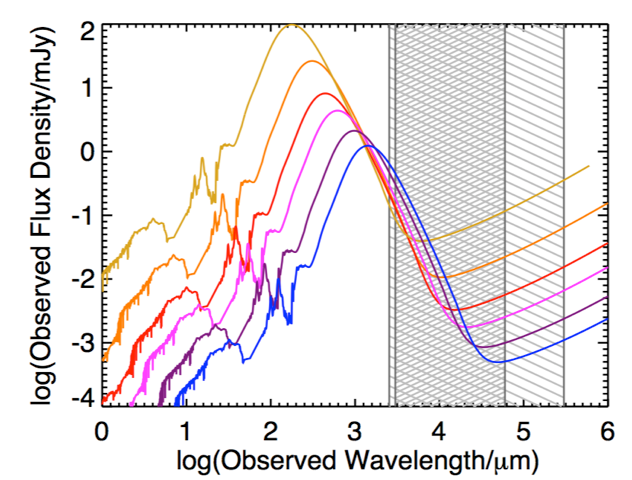}
\end{center}
\vskip -0.5cm
\begin{spacing}{0.7}
\caption{\footnotesize Observed flux density vs. wavelength for 
the SED of a typical SMG at redshifts of $z=1$ (yellow), $z=2.5$
(orange), $z=4$ (red), $z=6$ (magenta), $z=10$ (purple), and $z=15$
(blue).  The gray regions indicate possible frequency coverage ranges
for the ngVLA of 5-100\,GHz and 1-118\,GHz, demonstrating that the
ngVLA will be able to sample the full radio spectral energy
distributions (SEDs) of galaxies out to high-redshift.  }
\end{spacing}
\label{fig:fullsed}
\end{figure}

\subsection{The promise of free-free as a SFR tracer}

Although lower frequencies have the advantage of a larger
field-of-view, which is necessary to get large numbers of very
high-redshift galaxies, there are several major issues to consider to
determine which frequencies would be best suited for mapping the star
formation history in the radio.  First, what are the dominant physical
mechanisms contributing to the radio emission at high redshifts?
Second, how well can we calibrate these mechanisms to estimate SFRs?
Third, can we separate AGNs from star-forming galaxies?  With
synchrotron emission, there is substantial contamination of the
star-forming samples by AGNs, and there are uncertainties in the
conversion from radio flux to star formation rate (SFR).

Free-free emission is dominant at rest-frame frequencies of tens of
GHz and is directly proportional to the production rate of ionizing
photons by young, massive stars (though it may be biased towards
starbursts), making it useful for measuring
SFRs \citep[e.g.][]{murphy15a}.  However, even at these high
frequencies there can be substantial synchrotron contributions that
have to be taken into account. For example, \citet{rabidoux14a} find a
free-free fraction of 55\%\ at 33~GHz for a sample of local
star-forming galaxies.  Using multiple measurements over the broad wavelength coverage of the ngVLA, we will be able to observe the full
radio spectral energy distributions (SEDs; Figure~\ref{fig:fullsed}) of galaxies out to high-redshift, 
allowing us to separate the pure synchrotron and free-free regions of the spectra and 
hence estimate the relative fractions produced by each mechanism as a function of frequency 
for determining more accurate SFRs.

\begin{figure}
\begin{center}
\includegraphics[natwidth=4in,width=4.0in]{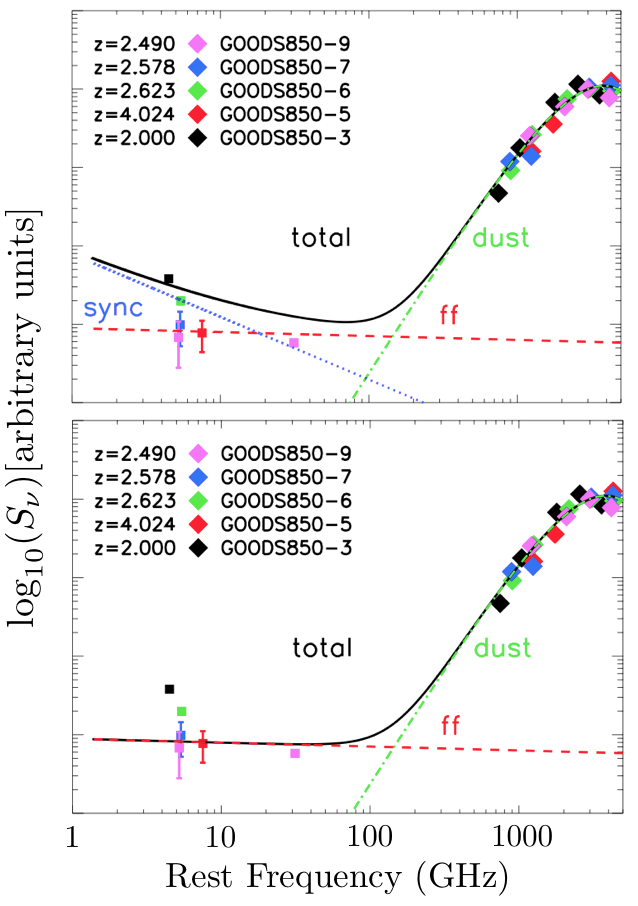}
\end{center}
\vskip -0.5cm
\begin{spacing}{0.7}
\caption{\footnotesize
Rest-frame flux density vs. rest-frame frequency for (a) either a moderate
luminosity, low-redshift galaxy or a ULIRG where the synchrotron
emission is not suppressed by the CMBR, and (b) a moderate luminosity,
high-redshift galaxy where the synchrotron emission is
suppressed. Data points show the SEDs of five isolated submillimeter
galaxies with redshifts from \citet{Barger12a}.  }
\end{spacing}
\label{fig:FF}
\end{figure}

One important factor to consider in this regard is the quenching of the synchrotron emission by the
cosmic microwave background radiation (CMBR).  Compton cooling of
relativistic electrons on the CMBR dominates over the synchrotron
emission when the CMBR energy density exceeds the galaxy magnetic
energy density \citep[e.g.][]{condon92a,murphy09a}.  Since the energy
density of the CMBR goes as $(1+z)^4$, there is a redshift
dependence. Moreover, the synchrotron emission will be highly quenched
in moderate SFR galaxies at high redshifts but possibly never in
ULIRGs, so there is also a luminosity dependence.  This means that
high-redshift galaxies with moderate SFRs will be totally free-free,
modulo the possible contribution from anomalous dust
emission \citep{adam15a}, which may arise from spinning dust grains
\citep[e.g.][]{draine98a}.  We illustrate the various cases in
Figure~\ref{fig:FF}.  In (a) we show the synchrotron emission (dotted;
assuming a constant spectral index) as it would be for either a
moderate luminosity, low-redshift galaxy or a ULIRG, as well as the
free-free emission (dashed), the dust emission (dot-dashed), and the
total emission (solid), while in (b) we show the synchrotron emission
suppressed as it would be for a moderate luminosity, high-redshift
galaxy.  Thus, the most secure SFRs will come from observations made
at high frequencies (rest-frame $30-50$\,GHz) where the free-free
emission significantly dominates for all galaxy types.  This range
will be readily accessible out to high-redshifts with the ngVLA.

\begin{figure}
\begin{center}
\includegraphics[width=4.0in,natwidth=4in]{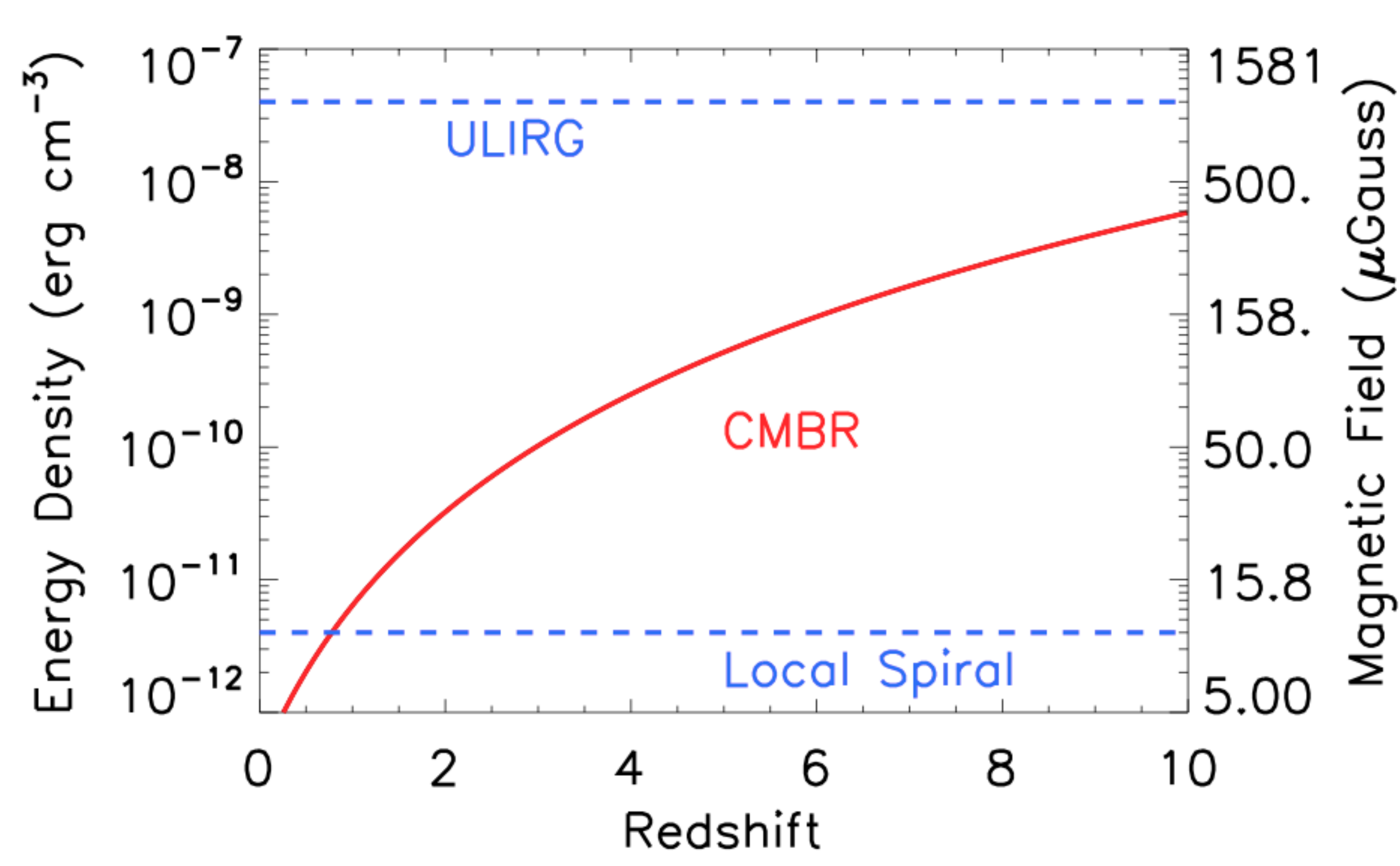}
\end{center}
\vskip -0.5cm
\begin{spacing}{0.7}
\caption{\footnotesize
Energy density in ambient photon field and magnetic field vs. redshift for a ULIRG and a local
spiral \citep{carilli01,mcbride14a}, and for the redshift-dependent CMBR.  The
corresponding magnetic field strength, assuming equipartition, is
shown on the right-hand axis.}
\end{spacing}
\label{fig:CMBR}
\end{figure}

One potentially interesting benefit of the CMBR quenching of the
synchrotron emission is that by mapping out the radio SEDs for
a number of galaxies of a given luminosity with the ngVLA and seeing
at what redshift the shape changes from synchrotron plus free-free to
free-free only, we can get a rough galactic magnetic field measurement
for that luminosity by equating it to the CMBR energy density \citep[e.g.][]{carilli01}. We show
this in Figure~\ref{fig:CMBR}, where we plot energy density versus redshift for a
local spiral and a ULIRG \citep{mcbride14a}, and for the
redshift-dependent CMBR.  We show the magnetic field strength on the
right-hand vertical axis.

Another benefit of the CMBR quenching is that extended AGNs are
expected to be quenched at high redshifts since the magnetic fields
in the extended radio jets are small, leaving only compact
AGNs \citep[e.g.][]{afonso14a}.  Since the typical radio
sizes for star-forming galaxies ($\sim1''$) are larger than those for
compact AGNs (Cowie \etal, in prep), it should be straightforward to
separate star-forming galaxies from AGNs even with only moderately
sub-arcsecond spatial resolution.

\subsection{Thermal dust emission at the highest redshifts}

Thermal re-radiation from dust dominates a galaxy's spectrum at rest
frequencies above $\sim$200\,GHz. This dust emission is dominantly
powered by recently-formed, massive stars, making it an excellent
tracer of the bolometric luminosity in dusty starbursts. While ALMA is
already revolutionizing our understanding of dusty star forming
galaxies at intermediate redshifts \citep[e.g.,][]{Karim13a,hodge13a,
Weiss13,Hezaveh13a,Simpson13a}, at sufficiently high-redshift, the
ngVLA bands will also benefit from the very-negative K-correction on
the cold dust Raleigh-Jeans tail.  As a consequence, the higher
frequency coverage of the ngVLA will bring with it the exciting
possibility of detecting thermal emission from dust at the highest
($z\sim7$) redshifts, providing important constraints on high-$z$ dust
emission.  

One crucial caveat is that detecting this thermal dust emission against the background of the CMB becomes more difficult at higher redshift \citep{dacunha13a}. As discussed in Section~\ref{sec:deepfield} in regards to CO emission, the net effect of the higher CMB temperature at high-z is to decrease the contrast between the target and CMB, 
implying that detecting thermal dust emission at the highest redshifts may only be feasible for the most luminous star-formers.
It will be critical to take this effect into account even for \textit{detected} galaxies, as the contrast against the CMB is frequency dependent. Failing to account for this effect could therefore affect the interpretation of the dust SED, yielding biased estimates of dust temperature, mass, emissivity index and luminosity \citep{dacunha13a}.

\section{AGN and supermassive black holes}

The ngVLA will have a large impact on studies of the evolution and
characteristics of massive black holes at the centers of distant
galaxies. In this section we outline some key science goals related to
the evolution of active galactic nuclei (AGN) and how the ngVLA will
be uniquely useful.

\subsection{Black hole masses from gas dynamics}

The proposed capabilities of the ngVLA offer the tantalizing
possibility of directly observing gas dynamics in the vicinity of the
super-massive black holes at the centers of massive galaxies, and
thereby also of measuring the masses not just of AGN, but of quiescent
super-massive black holes. Such observations require being able to
resolve the ``radius of influence'' of a black hole, i.e. the distance
within which the black hole's gravitational potential significantly
affects the orbital motion of the surrounding stars and interstellar
gas. From \citet{krolik99a}, the radius of influence is given
approximately by $r = (M_{\rm BH}/M_{\odot}) (\sigma/200 {\rm km/s})
^{-2} $. From the M-$\sigma$ relation as calibrated by \citet{kormendy13a}

 \[ {\rm log_{10}} ( \sigma/200 {\rm km/s}) \sim 0.23 ( {\rm
  log_{10}} M_{\rm BH} - 8.5 ). \] Combining these two formulae yields
\[ r \sim 7.5\times 10^{-4} (M_{\rm BH}/M_{\odot})^{(0.54)} {\rm ~pc} \]

Thus, a $10^9$\,\msun\ black hole has a radius of influence of about
54\,pc, corresponding to almost 30~mas at $z=0.1$, easily resolvable
to the ngVLA with long baselines. The ngVLA will revolutionize
estimates of black hole masses in the large fraction of systems that
are heavily obscured in the optical/near-IR, where estimates from
optical gas or stellar dynamics cannot be obtained even with 30\,m
class optical/near-IR telescopes like GMT, TMT, and the E-ELT.

Observations from the ngVLA will also enable high resolution
observations of galaxy dynamics (see discussion in
\S~\ref{sec:dynamics}) to constrain models that can be used to
estimate velocity dispersions for use with the $M-\sigma$ relation in
high-$z$ galaxies. Current observations barely resolve the host
galaxies, making reliable separation of circular velocity and velocity
dispersion difficult \citep[e.g.][]{kimball15a}.  Although resolving
the dynamic across parsec scales may be limited in practice to the
$z\simlt0.1$, full galaxy wide dynamics will be essential for
characterizing distant sources on the $M-\sigma$ relation.

\subsection{The nature of radio emission from radio-quiet AGN}

Radio emission from radio-loud AGN is powered by synchroton emission
from relativistic jets emanating from a region very close to the
central black hole. However, even traditionally radio-quiet objects
have significant radio emission, the origin of which is unclear. The
radio emission may originate in scaled down versions of the jets of
radio-loud objects, in synchrotron emission from material shocked by
thermal winds from the accretion disk \citep{zakamska14a}, or from
star formation in the quasar host \citep[e.g.][]{kimball11a}.
Determining which of these dominates is important, especially for our
understanding of the feedback process in radio-quiet objects.  If jets
are ubiquitous, they can be agents of feedback along with any thermal
winds. If star formation dominates, it implies that most quasar hosts
are heavily star-forming at the level of ULIRGs, which would be a
challenge for models in which AGN feedback is supposed to inhibit star
formation.

Determining which process dominates requires high spatial resolution,
multifrequency observations to distinguish the jets and flat-spectrum
cores expected from weak radio AGN from more diffuse steep spectrum
emission from thermal shocks, and a combination of synchrotron and
free-free thermal emission (at high frequencies) from star formation.

\subsection{Weak radio AGN in normal galaxies}

Our own galaxy contains a weak radio AGN, SgrA$^{*}$, with a
luminosity in its flat spectrum core $L_{8GHz} \sim 10^{16} {\rm
  WHz^{-1}}$.  The ngVLA could detect such an AGN in nearby galaxies
out to $\approx 8$\,Mpc with a flux density of 1\,\uJy\ (enclosing the
nearest $\sim 100$ nearby galaxies), with a $10\sigma$ detection in
10\,hours.  A more luminous radio AGN, such as that in M87
($L_{8GHz}\sim 10^{24}{\rm WHz^{-1}}$) could even be seen out to
$z>6$. Such AGN are important for ``maintenance'' or ``radio'' mode
feedback in massive galaxies, and seem only to be produced by the most
massive black holes. They can thus trace the evolution of these black
holes from when they form at $z\sim5$ to the current epoch. Although
the SKA will excel at finding many millions of such objects,
isolating those at high redshift will require additional selection
using high frequency data.  In these data we can pinpoint the steepest
spectrum objects, which the ngVLA will easily identify. The ngVLA will
also be the only telescope capable of determining the morphologies of
these objects and studying how the radio jets interact with the ISM of
their host galaxies.

\subsection{Multiple AGN}

The combination of high resolution and high sensitivity will allow the
ngVLA to resolve two merging AGN within their host galaxies. At
10\,GHz, the ngVLA's ``goal'' resolution is 30\,mas, which could
easily distinguish two AGN separated by $\sim 300$pc at $z\sim 1$.
Thus, the fraction of dual and multiple AGN can be used to constrain
black hole merger rates \citep[e.g.][]{fu15a} and the contribution of
mergers to the growth of supermassive black holes.

If a VLBI component with intercontinental baselines is added to the ngVLA, 
this opens up the possibility of obtaining baselines with high sensitivity 
at pc-scale resolutions for AGN at $z\sim 0.1$. This is important, as, even after decades
of work, it is still unclear how, or even if, the final stages of supermassive black hole
mergers take place in the centers of galaxies. This {\it final parsec problem} refers
to the difficulty of understanding how a binary black hole loses angular momentum
in a regime where neither dynamical friction with other stars nor gravitational radiation
is effective. Although various theories have been put forward to overcome this
problem \citep[e.g.][]{vas15}, statistics of binary black hole separations in 
galaxy cores with a range of dynamical properties are needed to constrain these models.

\subsection{Molecular gas and feedback in AGN}

In order to understand the effects of AGN (and stellar) feedback on
galaxy formation, it is important to be able to study molecular
outflows, not just in CO, but also in dense gas tracers (HCN, HCO+,
CS, etc, as discussed in \S~\ref{sec:densegas}) where CO is optically
thick. It remains an open question as to whether the most important
feedback mechanism is through expulsion of gas during short, but
violent feedback episodes in ``quasar'' mode, or ``radio'' mode, where
little, if any outflowing gas exceeds the escape velocity of the
system, but jets can inject turbulence into the ISM sufficient to
prevent star formation.  The ngVLA will allow the detailed study of
gas dynamics in the host galaxy to help us answer this important
question.

Another interesting application of the ngVLA will be the study of gas
chemistry in AGN versus star-forming galaxies. A highly obscured AGN
is very difficult to find using any conventional technique: optical
line diagnostics, X-rays and mid-IR emission from hot dust can all be
blocked by a sufficient column of dust and gas; far-infrared emission
is hard to distinguish from star formation; and radio free-free
emission can arise either from AGN photoionization or from star
formation. It has been speculated, however, that some molecular
species may be more (or less) common in regions dominated by XDR
chemistry, due to an AGN, or PDR chemistry, due to star formation
\citep[e.g.][]{martin15a,izumi15}.  If so, detection of molecular transitions
in the 3\,mm band (100\,GHz) of ngVLA will provide the best evidence
possible of the presence of very highly obscured AGN, and some of the
lower-J transitions at lower frequencies.

\subsection{Sunyaev-Zeldovich Effect from Quasar/Starburst Winds}

Thermal winds from AGN and starbursts are predicted to be detectable
via the Sunyaev-Zeldovich effect in long (several tens to hundreds of
hours) integrations using ALMA \citep{chatterjee08,rowe11}, an ngVLA
would be many times faster at detecting these, and could be used to
characterize the outflows in extent and pressure, constraining quasar
lifetimes and providing important information to feedback
models. There has been a claim of a statistical detection of quasar
winds via this technique in Planck data \citep{ruan15}, but only the
ngVLA will be able to study it in detail in individual objects.



\section{Polarimetry and Cosmic Magnetism}

Magnetic fields are ubiquitous in nature and have been detected on all astrophysical
scales up to, and tentatively beyond, those of galaxy clusters. Magnetic fields
can be dynamically important, for example having comparable energy density to
gravity and turbulence in interstellar gas \citep{beck07a}, and are
essential for understanding a wealth of astrophysical processes from star to
galaxy formation. The buildup and impact of magnetic fields over cosmic time
remain outstanding problems in astrophysics. Polarimetry is a powerful tool for
unveiling the magnetic Universe, and for providing a window into associated
science such as radiative transfer theory, properties of dust, and black hole
accretion flows.

To date, far-IR and millimeter polarimetry with facilities such as the
JCMT and CARMA have been limited to angular resolutions of a few arcseconds.
Sub-arcsecond resolution can be attained with the VLA and ALMA, and also with
the VLBA albeit at limited surface brightness sensitivity. Polarimetry with
ALMA and the recently upgraded Jansky VLA is only just beginning and will
soon deliver scientific yields. The groundwork from these studies will set the
stage for research directions within the new parameter space of the ngVLA.
Key to this new parameter space is angular resolution, which will
minimize beam depolarization (vector averaging of polarization position angles
within a large observing beam), enabling quantitative studies of magnetic
fields in unprecedented detail. The sensitivity improvement of the ngVLA will
enable the lowest signal-to-noise total intensity science with the VLA and ALMA
to be studied in polarization.

The SKA will play a fundamental role in tracing cosmic magnetism through studies of synchrotron radiation and Faraday rotation. By going to higher frequencies, the ngVLA will open a unique window on cosmic magnetic fields and science accessible through microwave polarimetry. The 5-100 GHz frequency range proposed for the ngVLA will facilitate observations of polarized emission from dust, atomic and molecular spectral lines, and synchrotron radiation from extreme Faraday rotation environments as well as those that are significant depolarized at lower frequencies. The ultra wide bandwidths proposed for the ngVLA will allow detailed Faraday structure mapping. We address each of these with example science themes in the following sections.
We note that while the following sections necessarily
focus on extragalactic science, much of the discussions are also relevant for
Working Groups 1 (Cradle of Life) and 2 (Galaxy Ecosystems) through topics
such as cometary comae, circumstellar disks, planetary nebulae, and Galactic
molecular clouds.

\subsection{Dust Continuum}

Linearly and circularly polarized thermal emission at far-IR and millimeter wavelengths
can be produced by elongated, spinning, aligned dust grains. Despite over half a
century of effort, a comprehensive theory does not yet exist to explain exactly how
dust grains are aligned \citep{Lazarian07a}. Observations indicate that
alignment is with respect to the magnetic field, though the alignment mechanism may
not be magnetic. Promising candidates include radiative alignment via torques that are
efficient when grain sizes are comparable to the wavelength, and mechanical alignment
in the presence of gas flows. The degree of fractional polarization can
provide diagnostics of grain characteristics \citep{hildebrand95a},
Typical observed fractions in molecular clouds are a few percent
\citep{leach91a,greaves99a}. When measurements of turbulent
velocities and the dispersion of polarization position angles across a source
are available, the Chandrasekhar-Fermi method \citep{chandrasekhar53a} can be
used to estimate magnetic field strengths in the plane of the sky
\citep[e.g.][]{falceta-goncalves08a}. This method requires the dispersion of
position angles to be meaningfully coupled to turbulent structure. If
this condition is not met, for example due to beam depolarization,
then magnetic field strengths will be
overestimated \citep{houde09a}. Similarly, not all tangled magnetic
field structures will be sampled along the line of sight (LOS) when the
emission is optically thin, leading to overestimated magnetic field
strengths. The three-dimensional magnetic field distribution can be
reconstructed through a combination of dust polarimetry and
measurements of ion-to-neutral molecular line width
ratios \citep{houde05a}.

The ngVLA will be capable of mapping magnetic fields in the molecular
environments of a large number of nearby galaxies through dust
polarimetry. For example, an ngVLA with collecting area 5~x~VLA could
map the millimeter wavelength dust polarization and magnetic field
geometry in a galaxy like the starburst M82
\citep[$d=3.5$~Mpc;][]{greaves00a,reissl14a} in less than 10h
if it were located at a distance of 350~Mpc, i.e. well beyond the
boundary of the Local Supercluster (Laniakea), sampling neighboring
superclusters. Such studies will be highly complementary to other
ngVLA probes of the interstellar medium in galaxies such as
(polarized) synchrotron emission and molecular line diagnostics, and
also to lower frequency studies of nearby galaxies and their magnetism
with the SKA \citep{beck15a}.  This will enable a detailed
understanding of the environments of local galaxies, including the
interaction between gas and magnetic fields, in turn providing
constraints on the end-points ($z=0$) of magnetized galaxy evolution
models
\citep[e.g.][]{van-eck15a}. Studies of local volume supernova remnants
through their sometimes highly fractionally polarized dust emission
\citep{dunne09a} may be developed to similar effect. ngVLA studies of
dust polarization in environments where grain alignment is expected to
be occurring, such as accretion disks \citep{aitken02a}, will enable
new probes of magnetic fields.


\subsection{Spectral Lines}

Rotational emission lines of molecules can be linearly polarized
through the Goldreich-Kylafis (G-K) effect \citep{goldreich82a}. This
arises when excited states of a molecule are exposed to anisotropic
radiation, either from an external source (e.g. nearby infrared
emitter) or internally from a gradient in the line optical depth
(e.g. due to a local velocity gradient). De-excitation then results in
spectral lines with net linear polarization, with fractional levels
typically a few percent. The production of linear polarization in
absorption lines against unpolarized continuum sources, as well as the
production of both linear and circular polarization in the absorption
lines against linearly polarized continuum sources, will also arise in
the presence of optical depth
anisotropies \citep{kylafis83a}.

By utilizing the G-K effect, the ngVLA will be capable of mapping
magnetic fields in the molecular environments of a large sample of
galaxies extending to high redshifts. 
For example, assuming a channel rms noise of 10 $\mu$Jy beam$^{-1}$, the ngVLA could detect CO(1--0) in a M$_{\rm gas}$ $=$ 10$^{10}$ M$_{\odot}$ galaxy out to $z\sim0.85$ with sufficient sensitivity to detect a few percent polarization.
Recently, \citet{li11a} mapped
CO polarization in M33 to investigate the interplay between large and
small scale magnetic fields on the formation of molecular
clouds. ngVLA studies capturing more detail within larger source
samples will be critical for understanding the environmental impacts
of magnetic fields within galaxies over cosmic time.  Spectral line
polarimetry with the ngVLA may be essential for studying galactic
magnetic fields at high redshift, for example in main sequence
galaxies where synchrotron probes will be of limited use due to CMBR
quenching (see Figure~\ref{fig:CMBR}). 

The G-K effect will lead to polarization of polycyclic aromatic
hydrocarbon (PAH) emission. ngVLA studies of polarized anomalous
microwave emission \citep[AME;][]{draine98a} may therefore provide
insight into grain alignment theory and perhaps even the very nature
of PAHs. While such studies may be most productive within the Galaxy,
extragalactic investigations of AME \citep[e.g.][]{hensley14a} and its
polarization characteristics will offer important consistency checks.

Elliptical polarization of atomic and molecular emission and
absorption lines is produced by the Zeeman effect. The presence of a
magnetic field causes the spectral line to be split into three
components: a linearly polarized component that is unshifted in
frequency and two elliptically polarized components shifted
symmetrically above and below the original line frequency.  The
elliptical polarization is a combination of (intrinsic) circular
polarization proportional to the strength of the line of sight
magnetic field with opposite sign for each split component, and
(propagation induced) linear polarization proportional to the strength
of the magnetic field oriented perpendicular to the position angle in
the plane of the sky.  The magnitude of the frequency shift depends on
the strength of the total magnetic field and the splitting factor
(Land\'{e} g-factor), the latter dependent on the magnetic moment of
the species.  For most spectral lines except those associated with
strong masers (e.g. OH), the magnitude of the frequency splitting is
much less than the spectral line width, in which case the Zeeman
observations are only sensitive to the line of sight magnetic
component. Zeeman splitting is approximately 3 orders of magnitude
weaker for non-paramagnetic species than paramagnetic species.

The ngVLA frequency range will enable Zeeman studies of paramagnetic
molecules such as C$_4$H, SO, and C$_2$H (and potentially CN depending
on the upper frequency bound), non-paramagnetic molecules such as
H$_2$O and NH$_3$, and radio recombination lines from atomic species
such as H and C. Unsampled magnetic structures within the telescope
observing beam or along the line of sight will reduce the measured
Zeeman effect. The high angular resolution of the ngVLA is therefore
critical for maximizing Zeeman detectability, while its sensitivity
will be necessary to improve detection statistics in weak magnetic
environments or for species with low magnetic
moments. \citet{robishaw15a} present a selection of Galactic and
extragalactic science that can be addressed by SKA observations of the
Zeeman effect. 
Much of this is applicable to higher frequency science with the ngVLA,
for example mapping magnetic fields in galaxies using masers
or megamasers to constrain the redshift evolution and dynamics of
galactic magnetism.
\citet{robishaw15a} point out that measurements of the Zeeman effect
are currently sensitivity limited. The ngVLA can uniquely contribute
to this field.

Non-Zeeman circular polarization of molecular rotational spectral
lines, with fractional levels of a few percent, can arise from
resonant scattering through the conversion of linear to circular
polarization \citep{houde13a}.  This effect is proportional to the
square of the magnetic field component in the plane of the sky. The
effect could be examined in ngVLA data to improve the analysis of
Zeeman observations, or exploited for Zeeman-insensitive molecules.

\subsection{Synchrotron Continuum}

The synchrotron mechanism produces linearly polarized radiation with a
theoretical maximum fractional level of approximately 70\%\ in a
uniform magnetic field.  However, observed fractions at radio
wavelengths are often only a few percent due to selection effects such
as spectral (bandwidth) and spatial (beam) depolarization, depth
depolarization (Faraday dispersion), and the degree to which the
magnetic fields in the synchrotron emitting volumes are
ordered. Faraday rotation describes the rotation of the plane of
linear polarization as linearly polarized radiation passes through a
birefringent medium, such as a magnetized thermal gas containing free
electrons. The degree of rotation at wavelength $\lambda$ is
proportional to both $\lambda^2$ and the `rotation measure' (RM).
Observations of the position angle of linear polarization at multiple
$\lambda$'s enable measurement of the RM, which in turn is
proportional to the path integral of the line of sight magnetic field
component weighted by the electron density: RM~$\propto$~$\int
B_\textrm{\tiny LOS} \, n_e \,dl$. If $n_e$ can be measured or modeled
along the line of sight, then the magnetic field can be probed.

Faraday rotation studies will feature prominently with the SKA as part of the
cosmic magnetism key science driver. The higher frequency coverage of the
ngVLA will provide unique parameter space for studies of Faraday rotation,
with particular relevance for sources with very large rotation measures
({\footnotesize $\gtrsim$}~$10^4$~rad~m$^{-2}$) and those exhibiting
significant depolarization at lower frequencies. This is important,
for example, for studying accretion rates onto nearby supermassive
black holes such as Sgr~A* \citep{macquart06a} and M81*
\citep{brunthaler06a}. These sources exhibit large rotation measures
and are both depolarized at frequencies less than approximately
80~GHz.  The ngVLA can facilitate an accretion rate census of similar
nearby black holes.  For example, an ngVLA with collecting area
5~x~VLA could detect a black hole with similar millimeter wavelength
properties to Sgr~A* out to a distance of 1.5~Mpc in less than 10h,
i.e. encompassing a significant fraction of the local group. The high
angular resolution and sensitivity of the ngVLA will aid in the
observation and interpretation of sub-parsec scale jet launching
environments near supermassive black holes, with implications for
precession and supermassive black hole binarity. For example,
millimeter wavelength polarization position angles for BL Lac appear
to match the structural position angle of the
jet \citep{stirling03a}.

More generally, synchrotron polarization studies with the ngVLA at its
lowest frequencies will contribute to investigations into the nature
of radio emission from radio-quiet AGN and weak radio AGN in normal
galaxies \citep[for example, comparing their fractional polarization
properties with other source populations][]{massardi13a}, high
rotation measure sources and their
environments \citep{broderick07a,marti-vidal15a}, polarized spectral
energy distributions \citep{homan12a,farnes14a}, high angular
resolution observations of rotation measure gradients and helical
magnetic fields in jets \citep{hovatta12a,gabuzda15a}, and turbulence
in galaxies through depolarization studies.

%
%

\section{Synergies with other facilities}

Here we describe how the ngVLA will complement and contrast with the
capabilities of other cutting-edge observatories.  As the 21$^{st}$
century is clearly an era of multi-facility use, and multiwavelength
studies, it is essential that ngVLA's discoveries in the area of
galaxy evolution complement, but do not duplicate, the range of
measurements made by other facilities.

\subsection{The Atacama Large Millimeter Array (ALMA)}

ALMA’s design is optimized for high frequencies in the millimeter and
submillimeter regime; it cannot be rivaled by ngVLA at wavelengths
$\stackrel{<}{_{\sim}}$1mm, in particular due to the contrast of
weather at the ALMA site and VLA site. However, ALMA has about half
the collecting area of the current VLA, so will be much less sensitive
than the ngVLA at mm wavelengths where its wavelength range
overlaps. For objects with large angular scales (for example, nearby
galaxies and SZ decrements in galaxy clusters), its smaller dishes and
compact array will remain valuable. At low frequencies, only ALMA's
Band-3 (84-116\,GHz) is currently available on the telescope; however,
by the advent of ngVLA, we expect the antennas to be outfitted with
Band-1 and Band-2 receivers, allowing observations as low as
30\,GHz. For objects near the celestial equator, ALMA baselines could
therefore be used to fill in shorter baselines for ngVLA observations
of objects with structure on a wide range of angular scales.

Another area where we expect significant synergy between ALMA and the
ngVLA takes advantage of their very complementary angular
resolutions. The recent ALMA long baseline campaign showed that images
can be obtained with baselines out to 15\,km at 230\,GHz, providing a
resolution of 15\,mas. This resolution would match the ngVLA at
40\,GHz baseline length of 180\,km (5$\times$VLA). Working together,
the two telescopes would thus be sensitive to a wide range of CO
transitions (and the underlying dust continuum) at comparable angular
resolution, allowing studies of the spatially-resolved molecular gas
excitation and resolved star formation law in high redshift galaxies
\citep[e.g.,][]{hodge12a,hodge15a}.

\subsection{The Square Kilometre Array (SKA)}

The ngVLA will have a point-source sensitivity twice that of SKA
phase-1 (SKA-1 MID) in the frequency range where they overlap. The
upper frequency being considered for the SKA is between 14\,GHz and
24\,GHz. Though 24\,GHz and below could be useful for detecting
CO(1-0) at extremely high-redshift ($z > 4$), it falls short of the
frequencies where the impact of the ngVLA CO studies will be most
profound, namely, from 25--50\,GHz, covering low order CO (and other
molecular) transitions over the peak epoch of Cosmic star formation
($z \sim 1$ to 4).

In terms of brightness temperature sensitivity at 10GHz, a comparison
between the ngVLA specifications and \cite[]{murphy15b} suggests that
at $1^{‘’}$ resolution the surface brightness sensitivity in continuum
will be similar (RMS in 1hr with a 5GHz bandwidth $\approx 15$mK for
both SKA-1 MID and the tapered ngVLA beam), but at higher angular
resolutions ($\approx 30$mas) ngVLA will be about twice as sensitive.
Note that these conclusions depend ultimately on the adopted
configurations.  The strength of SKA for high redshift galaxy studies
is thus likely to be in HI and in continuum and polarization studies
that require high surface brightness sensitivity at GHz frequencies,
complementary to the ngVLA's focus on molecular gas and
higher-frequency thermal radio emission.
%
%

\subsection{Next Generation Bolometers on Single-Dish (Sub)mm telescopes}

Single-dish submillimeter and millimeter telescopes provide a unique
data product that submillimeter interferometric datasets cannot
provide: they can map large fields of view in multiple bandpasses
swiftly and simultaneously.  The {\it Herschel Space Observatory} has
contributed significantly to the last five years developments in large
field of view mapping, but is limited in scope due to its wavelength
range (the SPIRE instrument operated at 250--500$\mu$m), and is
particularly insensitive to the $z>2$ Universe.  The SCUBA-2
instrument in operation at the JCMT simultaneously maps at 450$\mu$m\ and
850$\mu$m, and is currently surveying large areas, up to 4\,deg$^2$ to
the 850$\mu$m\ confusion limit ($\sim$0.5\,mJy/beam).  These deep
surveys, and those at longer wavelengths to be done by bolometer
arrays like GISMO-2 at 1.2\,mm and 2\,mm, are ideal for probing DSFGs
in the $z>2$ Universe, and would be ideal candidates for ngVLA
spectroscopic follow-up.  This is particularly relevant since,
historically, redshift confirmation has been most challenging for
DSFGs at $z>3$, requiring direct confirmation through
far-infrared/millimeter wavelength emission lines.  The next
generation of wide-field submillimeter bolometers will lean on
facilities like the ngVLA significantly for quick and efficient
redshift follow-up.

\subsection{Future infrared astronomy space missions}

Two infrared space telescopes have been proposed for launch in the
next 1-2 decades. SPICA, a joint project of the Japanese and European
Space Agencies, will be a 3-m class cooled telescope, whose focus will
be on spectroscopic surveys of distant galaxies. SPICA is currently
scheduled for launch in 2025.  At the same time, NASA is currently
considering a proposal for a Far-Infrared Surveyor telescope, which
will be a 5-m class single-dish telescope, also concentrating on
spectroscopic surveys. Both SPICA and the Far-Infrared Surveyor will
be able to provide many targets for ngVLA follow-up of low order CO
lines and dense gas tracers, essential for understanding their
underlying gas mass.

\subsection{Large optical/IR telescopes (GMT, TMT, and E-ELT)} 

The ngVLA will have an angular resolution comparable to
adaptive-optics corrected images from the GMT, TMT or E-ELT
($\sim$15-20\,mas). This will allow, for example, direct morphological
comparisons of the ionized atomic and molecular gas in high redshift
galaxies on sub-kpc scales. Moreover, the combination of the ngVLA and
these large optical/IR telescopes will also allow dynamical
comparisons of these key ISM components, providing detailed insight
into the process of galaxy formation as a function of cosmic time.
This is particularly crucial for the formation of early, massive
galaxies, where star-formation rates would have to have been high, and
the dust content proportionally high.  By providing a direct route to
observing the star-forming molecular gas in high-redshift galaxies,
ngVLA observations will be necessary to interpret the complex
rest-frame optical and UV emission in the earliest galaxies.

\subsection{JWST}

JWST will be highly sensitive to emission from starlight in high
redshift galaxies up to at least $z\sim10$, and by the advent of the
ngVLA, JWST deep fields should be available for follow-up with the
ngVLA. A corollary of this synergy is that an ngVLA with survey
capability (for example, if fitted with phased array feeds) could be
used to perform a high frequency survey for CO-emitting galaxies in
these fields, some of which may not be visible -- even to JWST -- if
they are sufficiently dusty and/or have very low stellar mass.

\section{Summary, Outlook, \& Technical Requirements}

The ngVLA will be a truly unique facility, primed to reveal many of
the more elusive mysteries of galaxy formation and evolution.  In this
white paper, we have discussed some of the possible focus areas for
such a facility to the best of our ability and given the constraints
of current knowledge.  Weighing the science cases for these focus
areas, we identify five \textit{Key Science Projects} where a ngVLA
could have a significant impact on studies of galaxy assembly through
cosmic time, and which we summarize here:
\begin{enumerate}
\item\textbf{The cold gas history of the universe} \\[0.5mm]
To-date, our knowledge of galaxy assembly at high-z has been based
almost entirely on the stars and ionized gas which result from the
star formation process. Due to limitations in instrumentation, the
cold gas reservoirs that fuel the star formation have largely been
ignored. By blindly detecting the low-J CO emission in multiple
galaxies at once, a ngVLA would supply this crucial missing piece of
the galaxy assembly puzzle. Excellent sensitivity on relatively short
baselines (few km or less) could be critical to avoid over-resolution
in unbiased searches.

\item\textbf{Galactic dynamics at high-z} \\[0.5mm]
Even with hundreds of hours of telescope time on some of the brightest
galaxies in the universe, it remains challenging to resolve the cold
gas at high-z with the current VLA. The relative importance of mergers
and isolated disks in the early dust-enshrouded universe is thus an
open question. By easily resolving the cold gas reservoirs in high-z
galaxies on scales of a few hundred parsec to 1\,kpc, a ngVLA would
allow the community to characterize the dynamics of even the dustiest
systems at high-z.

\item\textbf{Tracing star formation in early galaxies with dense gas} \\[0.5mm]
Studies in the local universe suggest that dense gas may be a more
fundamental tracer of star formation than the cold gas typically
studied in more distant galaxies. An order of magnitude fainter than
CO, such gas is beyond the reach of the VLA for all but a handful of
strongly lensed hyper-starbursting galaxies. A ngVLA would facilitate
the detection of low-J dense gas tracers at high-z, directly revealing
the star-forming gas in galaxies during the peak epoch of star
formation and beyond.

\item\textbf{Measuring dust-unbiased star formation with free-free emission} \\[0.5mm]
Free-free emission is both relatively unaffected by dust obscuration
and directly proportional to the production rate of ionizing photons
by young massive stars, making it a promising tracer of star formation
at high-z. Unfortunately, its faintness makes it extremely difficult
to detect with current instruments, and even when it is detected, it
is difficult to quantify contamination by a synchrotron component. A
ngVLA would deliver the full radio spectral energy distributions of
high-z galaxies, making it possible to isolate the free-free component
of the spectrum and thus determine more accurate star formation rates.

\item\textbf{Magnetic fields in galaxies} \\[0.5mm]
The role of magnetic fields in star formation and galaxy evolution are
still poorly understood. A ngVLA would have the sensitivity, angular
resolution, and unique microwave frequency range to produce a detailed
observational census of magnetic fields in a wide range of galaxies
throughout the Local Supercluster (Laniakea) on scales from
supermassive black hole accretion flows up to molecular clouds. Such
work would be highly complementary to lower-frequency studies of
magnetic fields on typically larger scales with the SKA.

\end{enumerate}

\noindent As the ngVLA is still in the initial technical
design phase, it is important for us to clearly state the
specifications required by these key science projects for the endeavor
as a whole to be successful.  We summarize those technical remarks
here:

\begin{itemize}
\item {\bf Sensitivity.} A minimum of 5 times the collecting area 
of the current VLA would be necessary for a substantial impact on
  extragalactic science.  This will dramatically improve our ability
  to transform high-$z$ single-source CO(1-0) follow-up to blank-field
  CO(1-0) detection campaigns, and bring the number of high-$z$
  CO(1-0) detections from a handful to several thousands.  This type
  of contribution is substantial. The vast majority of the
  extragalactic community currently focuses on galaxies' UV, optical
  and near-infrared emission.  This leap in instrumentation capability
  in the 1\,cm regime has the potential to refocus the community's
  energy on the often dominant and always critical mechanisms for
  galaxies' emission: their dust and gas.  We point the reader back to
  Figure~\ref{fig:sensitivity} for a simple depiction of the improved
  sensitivity limits of ngVLA over the current VLA and ALMA (see also
  NRAO ngVLA Memo No. 5, Carilli \etal, Figure 2).

\item {\bf Antennae Size.}  With its 25\,m antennae, the current
  VLA has a relatively small field of view.  Assuming equal collecting
  area and smaller antennae (12\,m), this could increase the field of
  view substantially, enabling larger area mosaicked maps at equal
  cost.  The field of view has substantial impact on the number of
  sources detectable in molecular gas per pointing, from $\sim$45
  sources in a single 18\,hour exposure (with 12\,m antennae) down to
  10 sources (with 25\,m antennae).  There is a similar impact on the
  number of galaxies we will be able to characterize in a large-field
  mosaic, ranging from $\sim$1500 galaxies across $2.5<z<9.5$ (12\,m)
  to $\sim$300 (25\,m) with a 700\,hour investment.  We direct the
  reader to Figure~\ref{fig:deepfield1} for an illustration of the
  antennae size on the yield of CO(1-0) emitters at high-redshift,
  assuming equal collecting area.

\item {\bf Bandwidth.} Wide bandwidth is a very useful component of
  high-$z$ science with the ngVLA.  Prior limitations restricted
  spectral line science of high-$z$ sources in the radio in previous
  generations, as frequencies were often too poorly constrained to
  waste time on multiple tunings.  Both a 3:1 and 5:1 RF bandwidth
  ratio would change the landscape of spectral-line surveys for high
  redshift galaxies, where so many of the challenges faced by the
  community are rooted in simply securing adequate redshifts.  We
  point the reader to Figures~\ref{fig:coladder}
  and~\ref{fig:densegas} for an illustration of which lines would be
  simultaneously accessible to wide-bandwidth observations in the
  10--50\,GHz range.  Note that the jump from the current JVLA 2:1
  bandwidth ratio to 3:1 is not a substantial change (it is equivalent
  to being able to detect CO(1-0) in the following redshift ranges:
  $1.5<z<4$ at 2:1 and $1.5<z<6.5$ at 3:1), however a wider bandwidth
  certainly would help the efficiency of high-redshift molecular gas
  deep field searches by capturing all relevant lines across a huge
  chunk of cosmic time.

\item {\bf Spatial Resolution.} The characteristic resolution needed
  to characterize the kinematics and dynamics in distant galaxies is
  $\sim$0.1$''$ and is primarily limited by the sensitivity limits per
  beam. There is no need for extra long baselines for high-redshift
  work, with the exception of projects focused on resolving the black
  hole radius of influence in more distant galaxies, at $\sim$30\,mas
  resolution.  We highlight Figure~\ref{fig:gn20} as an illustration
  of resolving dynamics in high-redshift galaxies on 0.1$''$ scales,
  and Figure~\ref{fig:codncc} as the possible future capabilities of
  ngVLA in detecting detailed sub-structure in intrinsically fainter
  systems.

\item {\bf Frequency Coverage.}  Aside from our continuum emission
  goals, which would ideally make use of the entire frequency coverage
  available to the ngVLA (Figure~\ref{fig:fullsed}), the sweet spot
  for high-redshift science falls in the 10--50\,GHz range, where
  spectral features (both low-J CO and dense gas tracers) are most
  prevalent beyond $z\approx2$.  Again, we draw the reader's attention
  to Figures~\ref{fig:coladder} and~\ref{fig:densegas} for
  illustrations.
  
  \item {\bf Polarization Purity.} Low instrumental leakage in both
  linear and circular polarization is crucial for magnetic field
  science with the ngVLA. Dust polarimetry, the Goldreich-Kylafis
  effect, and the Zeeman effect require sensitivity to sub-percent
  polarized radiation. This, in turn, requires stable on-axis
  polarization with leakage terms small enough to enable beam-center
  calibration to 0.1\% in Stokes Q, U, and V. Calibration residuals no
  worse than about 0.5\% at the primary beam half power radius are
  required to enable polarization mapping of extended sources, which
  is of particular importance for dust polarimetry.

\end{itemize}

The next few decades of technological development for astronomical
instrumentation will certainly lead to baffling discoveries, many of
which will be difficult or impossible to predict.  The next generation
VLA will be an essential tool towards this end, fulfilling a unique
niche untapped by any other radio facility at 10s of GHz.




\end{document}